\documentclass[traditabstract]{aa}
\usepackage{graphicx}
\usepackage[]{natbib}
\usepackage{enumerate}
\def\lunits{$\rm erg\,s^{-1}$~}
\def\funits{$\rm erg\,cm^{-2}\,s^{-1}$~}
\def\cunits{$\rm cm^{-2}~$}

\def\xmm{{\it XMM-Newton~}}
\def\chandra{{\it Chandra~}} 
\def\lxl6{{$\rm L_X/L_{6\,\mu m}~$}}

\begin{document}
\title{The XMM deep survey in the CDFS IV. Candidate Compton-thick AGN}

%   \subtitle{ }

  \titlerunning{Heavily obscured AGN in the CDFS}
    \authorrunning{I. Georgantopoulos et al.}

   \author{I. Georgantopoulos\inst{1,2},
                    A. Comastri \inst{1},
                    C. Vignali \inst{3,1},
                    P. Ranalli \inst{2,1}
                    E. Rovilos \inst{1,4}, 
                    K. Iwasawa \inst{5}, 
                    R. Gilli \inst{1}, 
                    N. Cappelluti \inst{1},
                   F.  Carrera \inst{6},
                    J. Fritz \inst{7},
                    M. Brusa \inst{3,1,10},
                    D. Elbaz \inst{8},
                    R. J. Mullaney \inst{8},
                    N. Castello-Mor \inst{6},
                    X. Barcons \inst{6}, 
                    P. Tozzi \inst{9},
                    I. Balestra \inst{10},
                    S. Falocco \inst{6}
                                         }

   \offprints{I. Georgantopoulos, \email{ioannis.georgantopoulos@oabo.inaf.it}}

   \institute{INAF-Osservatorio Astronomico di Bologna, Via Ranzani 1, Bologna, 40127, Italy \\
              \and 
              Institute of Astronomy \& Astrophysics, National Observatory of Athens, 
              Palaia Penteli, 15236, Athens, Greece \\
              \and
             Physics \& Astronomy Department, University of Bologna, Viale Berti Pichat 6/2, Bologna, 40127, Italy \\
               \and 
              Physics Department, University of Durham, South Road, Durham, DH1 7RH, United Kingdom \\
               \and   
     ICREA and Institut de Ci\'{e}ncies del COSMOS (ICC), Universitat de Barcelona (IEEC-UB), Mart\'{i} i Franqu\'{e}s 1, 08028, Barcelona, Spain\\ 
     \and
     Instituto de Fisica de Cantabria (CSIC-Universidad de Cantabria), 39005 Santander, Spain \\
       \and 
        INAF-Osservatorio Astronomico di Padova, Vicolo dell'Osservatorio 5, 35122, Padova, Italy \\
        \and
         Irfu/Service Astrophysique, CEA-Saclay, Orme des Merisiers, 91191 Gif-sur-Yvette Cedex, France  \\
                     \and                  
              INAF Osservatorio Astronomico di Trieste, via G.B. Tiepolo 11, I-34143, Trieste, Italy \\
              \and
              Max Planck f\"{u}r Extraterrestrische Physik, Karl Schwarzschild strasse, Garching, 85748, Germany 
                 }

   \date{Received ; accepted }

\abstract{The Chandra Deep Field is the region of the sky with the highest concentration of X-ray data available: 4Ms of \chandra and 3Ms 
 of \xmm data, allowing excellent quality spectra to be extracted even for faint sources.  We take advantage of this in order to compile a sample of heavily obscured  
   Active Galactic Nuclei (AGN) using X-ray spectroscopy. We select our sample among the 176 brightest \xmm  sources, searching for either a) flat X-ray spectra ($\Gamma<1.4$ at the 90\% confidence level)
   suggestive of a reflection dominated continuum or b) an absorption turn-over suggestive of a column density higher than $\approx 10^{24}$ \cunits. 
     We find a sample of nine candidate heavily obscured sources satisfying the above criteria.
      Four of these show statistically significant FeK$\alpha$ lines with large equivalent widths (three out of four have EW 
       consistent with 1 keV) suggesting that these are the most secure Compton-thick AGN candidates. Two of these sources are transmission dominated while the other two are most probably 
     reflection dominated Compton-thick AGN. Although this sample of four sources is by no means statistically complete, it represents the best example of Compton-thick 
      sources found at moderate-to-high redshift with three sources at z=1.2-1.5 and  one source at z=3.7. 
    Using  {\it Spitzer} and {\it Herschel} observations, we estimate with good accuracy the X-ray to mid-IR ($12\rm \mu m$) luminosity ratio of our sources. 
    These are well below the average AGN relation, independently suggesting that these four sources are heavily obscured. 
\keywords {X-rays: galaxies; Infrared: galaxies}}
   \maketitle
%
%________________________________________________________________

\section{Introduction} 
The nature of the X-ray background (XRB) has been contentious since
its discovery 50 years ago \citep{Giacconi1962}.  The \chandra mission
confirmed that the background is made up of the summed emission from
Active Galactic Nuclei \citep{Brandt2005}.  The resolved fraction of
the XRB is about $\sim$90 \% in the 0.5-2 keV and 2-5 keV bands
\citep{Alexander2003, Xue2011}.  Optical spectroscopic follow-up
observations show that the peak of the redshift distribution of these
sources is at $z\sim0.7-1$ \citep{Barger2003, Silverman2010}. At higher
energies, the limited sensitivity hampers the resolution of a large
fraction of the XRB.  Narrow energy-band source-stacking shows that
this fraction reduces to $\sim$60\% over 5-8 keV, and only $\sim$50\%
above 8 keV \citep{Worsley2005, Worsley2006, Xue2012, Moretti2012}.

Still, it is at high energies where the bulk of the XRB energy density
is produced.  The peak of the X-ray background at 20-30\,keV
\citep[e.g.][]{Frontera2007,Churazov2007,Moretti2009} can be
reproduced only by invoking a significant number of heavily obscured
and Compton-thick AGN.  Compton-thick AGN are those where the
absorbing column densities exceed $\rm 1.5\times 10^{24}$ \cunits and
thus the attenuation of X-rays by photoelectric absorption is enhanced
by the scattering on electrons \citep[for reviews on Compton-thick AGN
properties and surveys see][]{Comastri2004,Georgantopoulos2012}.
However, the exact density of heavily obscured AGN required by X-ray
background synthesis models remains still under dispute
\citep*{Gilli2007,Sazonov2008,Treister2009, Ballantyne2011,
  Akylas2012}. In particular, the intrinsic fraction of Compton-thick AGN 
   may vary from 15 to over 35\% \citep[see discussion in][]{Akylas2012}. 
   Additional evidence of a numerous Compton-thick
population comes from the directly measured space density of black
holes in the local Universe \citep[see][]{Soltan1982}. It is found
that this space density is a factor of 1.5-2 higher than predicted by
the X-ray luminosity function \citep{Marconi2004,Merloni2008},
although the exact number depends on the assumed efficiency in the
conversion of gravitational energy to radiation.
 
The advent of the {\it Integral} and {\it SWIFT} missions helped to constrain the 
 Compton-thick population in the local Universe. These missions explored the X-ray sky at energies
above 10\,keV probably providing the  most unbiased samples of
Compton-thick AGN over the whole sky. Owing to the limited imaging
capabilities of these missions (carrying coded-mask detectors), the flux limit probed is
 bright ($\rm f_{15-55 keV} \sim 10^{-11}$\,\funits), allowing only the detection of AGN at very low
redshifts. These ultra-hard surveys did not detect large numbers of Compton-thick sources
 \citep[e.g.][]{Ajello2008,Tueller2008,Paltani2008,Winter2009,Burlon2011,Malizia2012,Goulding2011}. The
fraction of Compton-thick AGN in these surveys does not exceed a few percent of
the total AGN population. In contrast, optical and mid-IR 
surveys yield Compton-thick AGN fractions of between 10\% and 20\%
\citep{Akylas2009,Brightman2011}.  However, even the ultra-hard surveys 
 may miss a fraction of Compton-thick 
  AGN, i.e. those  with $\rm N_H>2\times 10^{24}$ \cunits. 
  As \citet{Burlon2011} point out this may explain the relative scarcity of Compton-thick 
  AGN found in the ultra-hard {\it Integral} and {\it SWIFT} surveys. 

At higher redshifts, a number of efforts have been made to identify 
Compton-thick AGN. For example, \citet{Gilli2010}  
 provide samples of Compton-thick AGN at moderate redshifts ($z\sim1$)
  using optical spectroscopy and in particular the [NeV] emission line. 
  Searches for high-redshift Compton-thick AGN in the mid-IR also attracted much interest, mainly 
   because the absorbed radiation heats the surrounding material and is re-emitted 
    at IR wavelengths. The techniques that have been employed 
     include the detection of a high 24$\rm \mu m$ emission 
      relative to the optical emission \citep[e.g.][]{Fiore2008, Georgantopoulos2008, Fiore2009, Treister2009b, Eckart2010},
       and the presence of a low X-ray-to-mid-IR luminosity ratio \citep[see][]{Alexander2008,Georgantopoulos2011b,Alexander2011}
        Most recently \citet{Georgantopoulos2011} proposed the presence of deep silicate features in
        mid-IR {\it Spitzer} spectra, although \citet{Goulding2012} argue that the silicate absorption in these sources 
         is related to the host galaxies. 
         Nevertheless, the most unambiguous method for finding Compton-thick sources relies on X-ray spectroscopy. 
 In X-ray wavelengths, the most systematic attempts include these by 
 \citet{Tozzi2006, Georgantopoulos2007, Georgantopoulos2009} 
  and more recently \citet{Brightman2012} all in the \chandra deep fields.   
  A number of Compton-thick sources  have been individually 
   discussed in \citet{Norman2002}, \citet{Iwasawa2005}, \citet{Comastri2011}, \citet{Feruglio2011}, \citet{Gilli2011}
    \citet{Georgantopoulos2011b} and \citet{Iwasawa2012}.  In particular, \citet{Comastri2011},
provided the most direct X-ray spectroscopic evidence yet for the presence of
Compton-thick nuclei at high redshift, reliably identifying two Compton-thick
AGN at z=1.536 and z=3.700.

 Here, we attempt to find unambiguous examples 
  of Compton-thick  AGN at moderate redshifts, extending the work of \citet{Comastri2011}.
 The CDFS is one of the regions of the sky 
  with the largest accumulation of multi-wavelength data available. In particular, 
   it is the area with the most sensitive X-ray observations, namely 3 Ms of \xmm 
    data and 4Ms of \chandra data, allowing the extraction of good quality 
     X-ray spectra even for faint X-ray sources such as heavily obscured AGN. 
   Owing to the superior \xmm photon collecting power, we first select a sample of 
    candidate heavily obscured AGN using the 3Ms \xmm observations.
     For these sources, we present a combined \xmm and \chandra spectral analysis
      in order to increase the photon statistics.  
       Our aim is to detect unambiguous signs of Compton-thick
obscuration, such as direct detection of a large column density, or a flat
spectral index, and finally a large equivalent-width (hereafter EW) Fe\,K$\alpha$ line. 
 Finally, we examine the mid and far-IR (up to rest-frame wavelengths of 160$\mu m$)
   properties of our sources using data from the {\it Spitzer} 
  and {\it Herschel} missions. The aim is to check 
 whether the IR observations independently  support a heavy obscuration scenario. 
We adopt $\rm H_o=75\,km\,s^{-1}\,Mpc^{-1}$, $\rm\Omega_{M}=0.3$, and $\Omega_\Lambda=0.7$
throughout the paper.

\section{Data}

\subsection{X-ray}

\subsubsection{XMM-Newton}
The CDFS area was surveyed with \xmm during different epochs spread
over almost nine years. The data presented in this paper were obtained
by combining the observations awarded to our project and observed
between July 2008 and March 2010 \citep{Comastri2011} with the
archival data acquired in the period July 2001 - January 2002. The
total exposure time, after the removal of background flares, is
$\approx$2.82\,Ms for the two MOS and $\approx$2.45\,Ms for the PN
detectors.  The total area covered is 30$\times$35 arcmin. The source
catalogue contains 339 X-ray sources detected in the 2-10 keV band
with a significance larger than 5$\sigma$  and a flux density limit of $\approx
6.6\times10^{-16}$ \funits assuming $\Gamma=1.7$. A supplementary list
of 74 sources detected with lesser significance is also provided. An
extended and detailed description of the full data set, including data
analysis and reduction and the X-ray catalogue will be published in
 \citet{Ranalli2013}.

\subsubsection{Chandra}

The CDFS 4\,Ms observations consist of 53 pointings obtained in the years 2000
(1\,Ms), 2007 (1\,Ms), and 2010 (2\,Ms). The analysis of the first 1\,Ms data is
presented in \citet{Giacconi2002} and \citet{Alexander2003}, while the analysis
of the 23 observations obtained up to 2007 is presented in \citet{Luo2008}. 
 In the present 4Ms survey, 740 sources are detected down to a 
sensitivity limit of $\sim0.7\times 10^{-16}$\,\funits and
$1\times 10^{-17}$\,\funits in the hard (2-8\,keV) and soft (0.5-2\,keV) band,
respectively \citep{Xue2011}. The Galactic column density towards the CDFS is
$0.9\times 10^{20}$\,\cunits \citep{Dickey1990}.

\subsection{Infrared}

\subsubsection{Spitzer} 
The central regions of the CDFS were observed in the mid-IR by the
{\it Spitzer} mission as part of the Great Observatories Origins Deep Survey
(GOODS). These observations cover areas of about $10 \times 16.5\rm\,arcmin^2$
 using the  IRAC (3.6, 4.5, 5.8, and 8.0$\, \rm \mu m$)  bands. 
   These data are combined with more recent observations of the wider 
    E-CDFS area in the
SIMPLE survey \citep{Damen2011}. The combined data-set
has a 5$\sigma$ magnitude limit of [3.6 $\rm \mu m$]AB = 23.86, while the 3$\sigma$
magnitude limit of the central GOODS region is [3.6 $\rm \mu m$]AB =26.15.
The GOODS area in the centre of the CDFS has also been imaged
with {\it Spitzer}-MIPS in the 24 $\rm \mu m$ band with a 5$\sigma$  flux density
limit of 30 $\rm \mu$Jy. A much wider area, including the entire E-CDFS
was imaged as part of the FIDEL legacy program \citep{Magnelli2009} with a 5$\sigma$ 
 flux  density limit of 70 $\mu $Jy; we use a combination of the two data-sets for this work.

\subsubsection{Herschel} 
 The far-Infrared  data in this work come from the GOODS-Herschel survey of \citet{Elbaz2011}. 
 This is the deepest survey of {\it Herschel} using the PACS instrument in both the 100 and the 160 $\rm \mu m$ bands, 
  with an integration time of more than 15 hr per position. 
   The area covered is a 13$\times$ 11 arcmin field inside the GOODS area. The source detection 
    is performed using a 24$\rm \mu m$ prior position and the 3$\sigma$ flux density limits are 0.8 and 2.4 mJy in the 100 and 160
     $\rm \mu m$  bands respectively. 
    The GOODS-Herschel catalogue we use also utilises public data from the HerMES survey  \citep{Oliver2012} in the CDFS. 
    We use a  250$\rm \mu m$ catalogue based on 24$\rm \mu m$ prior positions covering the GOODS-S area, and we keep sources 
    with a flux density determination better than the 2$\sigma$ limit of 3.5mJy. 
   
\section{Sample definition and X-ray spectral extraction}

\subsection{The sample}
We confine our analysis to the brightest sources in the preliminary
\xmm catalogue i.e. those with a detection probability of
$8\sigma$. There are 194 sources from both the main and supplementary
lists satisfying the above criterion. However, for a number of sources
we cannot derive an \xmm spectrum in any of the cameras, either
because they are too faint (they lie in areas of elevated background
or low exposure) or they are confused. 
 The number of confused sources is determined using the \chandra imaging.
 Therefore, The number of extracted spectra
is 176.  We match the \xmm positions with the \chandra
positions using a radius of 5 arcsec, and find counterparts for all
the 176 sources, in the 4 Ms catalogue of \citet{Xue2011}, or the
E-CDFS catalogue of \citet{Lehmer2005}.

We build the multi-wavelength catalogue 
using all the information available and the likelihood ratio
method to select the counterparts, using the positional uncertainties
provided in the various catalogues. We first combine the
SIMPLE catalogue with both the K-selected (Taylor et al., 2009)
and the BVR-selected (Gawiser et al., 2006) MUSYC catalogues
(preferring K-selected sources in cases where they are detected
in both catalogues). We find the optical-infrared counterparts of
the X-ray sources, constraining their positions, and then we look
for counterparts in the FIDEL and 24 $\rm \mu m$-prior Herschel catalogues. 

Out of our 176 sources with \chandra counterparts, 136
have a spectroscopic redshift determination in \citet{Szokoly2004},
\citet{LeFevre2005}, \citet{Norris2006}, \citet{Ravikumar2007},
\citet{Vanzella2008}, \citet{Treister2009}, \citet{Balestra2010},
\citet{Silverman2010}, and \citet{Cooper2011}. For 38 of the remaining sources,
photometric redshifts have been compiled from \citet{Cardamone2010},
\citet{Taylor2009}, \citet{Rafferty2011}, and \citet{Luo2010}, \citet{Dahlen2010} while two sources
have no redshift determination, because they are too faint in optical --
near-Infrared wavelengths.

\subsection{XMM-Newton spectra} 
For each individual \xmm orbit, source counts were collected from
circular regions with radii between 10 and 25 arcsec, depending on
field crowdedness and source brightness, and centered on the 2-10 keV source
positions. These radii correspond to encircled energy fractions
of 59\% and 86\%, respectively considering the on-axis \xmm PSF at an energy of 4.2
keV (the average energy of a source with a spectrum of $\Gamma=1$).  
 The encircled energy fraction does not change abruptly with the off-axis angle or the average energy. 
  The above fractions become 55 and 80\% at an off-axis angle of 9 arcmin 
 and for an energy of 6keV. 
 Local background data were taken from nearby regions, separately
for the PN, MOS1, and MOS2 detectors to account for local background
variations and chip gaps, and avoiding \xmm or \chandra detected
sources.  The areas of the background regions on average have 20
arcsec radii.  The spectral data from individual exposures were summed
for the source and background, respectively, and a background
subtraction was made assuming a common scaling factor for the
source/background geometrical areas.  Both the PN and the MOS spectra
are extracted in the 0.5-8 keV energy area.  MOS1 and MOS2 spectra are
summed using the FTOOLS {\sl MATHPHA} task.  Response and effective
area files were computed by averaging the individual files using the
FTOOLS {\sl ADDRMF} and {\it ADDARF} tasks.

\subsection{Chandra spectra}
    We used the {\sc SPECEXTRACT} script in the {\sc CIAO} v4.2 software package to
extract the spectra of  \chandra sources. The extraction
radius varies between 2 and 4 arcsec with increasing off-axis angle. At low
off-axis angles ($<$4\,arcmin), this area encircles 90\% of the light at an energy of
1.5\,keV. The same script extracts response and auxiliary files. The addition
of the spectral files was performed with the
{\sc FTOOL} task {\sc MATHPHA}. To add  the response and auxiliary files, 
 we used the {\sc FTOOL}  {\sc ADDRMF}, and {\sc ADDARF} tasks respectively,
  weighting according to the number of photons in each spectrum.

%%%%%%%%%%%%%%%%%%%%%%%%%%%%%%%%%%%%%%%%%%%%%%%%%%%%%%

\begin{table*}
\centering
\caption{Initial selection of candidate Compton-thick sources based on \xmm absorbed power-law, ({\sc PLCABS}), spectral fits} 
\label{xmm} 
\begin{tabular}{cccccccccccccc}
\hline\hline 
 PID      &  LID &  XID & $z$    & $\rm N_H$     &  $\Gamma$  & EW &  c-stat  &  $\rm N_H$ &  Flux &   $\rm L_X^{obs}$  & $\rm L_X^{unobs}$  & Ref     \\
(1)    &   (2)    &  (3)   & (4)                 & (5)                & (6)   & (7)       & (8)    &            (9)       & (10)   & (11) & (12) & (13)  \\
\hline
48            &   334  & 198  & 0.298$^1$  &    $1.95^{+0.27}_{-0.19}$ & $1.23^{+0.04}_{-0.09}$   &  $0.07^{+0.05}_{-0.05}$ &  2909/2498  &   $3.3^{+0.17}_{-0.14}$ & 40.0 & 0.68  &  0.88 &     -  \\
66             &  126   & 191 & 1.185$^2$   & $12.3^{+45.0}_{-11.0}$ &     $-0.37^{+1.2}_{-0.6} $  &  $0.36^{+0.41}_{-0.31}$ &  3015/2498   & $100^{+22.0}_{-20.0}$ & 1.2      & 1.5    & 2.2 &   -    \\
144           &     265  & 412 & 3.700$^2$   &  $105^{+78.0}_{-65.0}$ &  $2.5^{+1.3}_{-1.4}$   & $0.40^{+1.80}_{-0.30}$  & 2845/2498   & $67.0^{+27.0}_{-21.0}$  &   1.55            & 3.7         &  40 & a,c,d,e    \\
147          &  176    & 257 & 1.537$^2$   &   $18.6^{+7.2}_{-7.7}$ & $0.43^{+0.59}_{-0.43}$     &  $0.51^{+0.30}_{-0.26} $ & 2881/2498   &       $61.0^{+9.0}_{-9.0} $ & 6.3       & 2.2    &   3.5 & a,b,d   \\ 
214          & 55 & 83 & 2.00$^3$  & $21.7^{+5.0}_{-7.1}$ &   $1.0^{+0.15}_{-0.25}$    & $0.21^{+0.09}_{-0.09}$ & 2874/2498 &    $48^{+2.5}_{-4.0}$ &  19 & 6.1  &  8.0 &  - \\
222           & 428 & 686 & 0.424$^2$ &  $0.30^{+0.37}_{-0.29}$ &  $0.77^{+0.24}_{-0.24}$ &  $<$0.27 & 2896/2498 & $2.0^{+0.52}_{-0.41}$ & 7.0 & 0.3 & 0.3 &  - \\
245         &     184 & 22  &  2.68$^3$ & $45.0^{+25.0}_{-15.0}$   & $0.93^{+0.38}_{-0.41}$  & $<$0.62   &  2820/2498 &  $91.0_{-22.6}^{+33.1}$ & 3.0 & 1.5 & 5.0 &  e   \\
289          &   128 & 193 & 0.607$^2$ &  $0.56^{+0.42}_{-0.21}$ & $0.82^{+0.15}_{-0.12}$ & $0.11^{+0.10}_{-0.10}$ & 2850/2498 & $3.2^{+0.4}_{-0.4}$ & 15.9 & 1.2 &  1.4 & -\\  
324          &  398   &  634 & 1.222$^2$ & $<0.63$ &    $0.69^{+0.30}_{-0.28}$    & $0.71^{+0.58}_{-0.45}$ & 2940/2498 & $6.0^{+6.8}_{-3.9}$ & 3.6 & 1.2 &  1.2 & f \\
\hline \hline 
\end{tabular}
\begin{list}{}{}
\item The columns are:
(1) \xmm  ID.
(2) \chandra ID from \citet{Luo2008}.  
(3) \chandra ID from \citet{Xue2011}. 
(4) redshift; two and three decimal digits denote  X-ray and spectroscopic redshift respectively; 
 reference: $^1$ \citet{Balestra2010}, $^2$ \citet{Szokoly2004},  $^3$ X-ray redshift from current work. 
(5) $\rm N_H$ column density in units of $\rm 10^{22}$ \cunits. 
(6) Photon index.  (7) FeK$\alpha$ equivalent-width in units of keV; the energy of the line has been fixed at a rest-frame energy 
 of 6.4 keV.
  (8) c-statistic value / degrees of freedom (9) Column density in units of $\rm 10^{22}$ \cunits 
  for a photon-index fixed at $\Gamma=1.8$. (10) 2-10  keV flux in units of 
 $10^{-15}$ \funits. (11) 2-10 keV obscured luminosity in units of $10^{43}$ \lunits. (12) Unobscured luminosity (estimated by removing the absorption component) 
  in units of $10^{43}$ \lunits.  The flux and luminosities are derived having $\Gamma$ free. 
  (13) previous X-ray reference:
  (a) \citet{Tozzi2006}; (b) \citet{Georgantopoulos2007}; (c) \citet{Norman2002}; (d) \citet{Comastri2011}; (e) \citet{Iwasawa2012};  (f) \citet{Georgantopoulos2011b}.  
   All errors refer to the 90\% confidence level. 
\end{list}
\end{table*} 
    
%%%%%%%%%%%%%%%%%%%%%%%%%%%%%%%%%%%%%%%%%%%%%%%%%%%%%%%%%%%%%%%%%%

\begin{table*}
\centering
\caption{Joint XMM and Chandra spectral fits of the full candidate Compton-thick sample} 
\label{xmm-chandra} 
\begin{tabular}{ccccccccccc}
\hline\hline 
\multicolumn{3}{c}{} & \multicolumn{4}{c}{Power-law} & \multicolumn{3}{c}{Reflection} \\
\hline
 PID  &  cts &  cts              & $z$    & $\rm N_H$     &  $\Gamma$  & EW & c-stat  &  $\Gamma$ & EW  & c-stat \\
(1)      &              (2)    &         (3)       & (4)                 & (5)                & (6)   & (7)       & (8)    &            (9)       & (10)  & (11)  \\
\hline
48           &   7736  & 4172 & 0.298  &    $2.51^{+0.23}_{-0.17}$ & $1.38^{+0.04}_{-0.09}$   &  $0.10^{+0.03}_{-0.03}$ & 3622/3010  & $2.55^{+0.03}_{-0.03}$ & $<$0.013  & 4078/3011         \\
66            &  1044   & 215 & 1.185   & $100^{+22.0}_{-19.0}$ &     $1.8$  &  $0.50^{+0.55}_{-0.28}$ &  3614/3010   &   1.8    & $0.55^{+0.30}_{-0.30}$    &  3666/3012        \\
144          &     848  & 407 & 3.700   &  $48.7^{+44.0}_{-27.0}$ &  $1.26^{+0.72}_{-0.47}$   & $0.61^{+0.43}_{-0.33}$  & 3477/3010   &   $1.91^{+0.19}_{-0.14}$  &  $0.564^{+0.30}_{-0.30}$     & 3477/3011       \\
147          &  1207    & 593 & 1.537   &   $18.6^{+5.2}_{-7.0}$ & $0.46^{+0.45}_{-0.36}$     &  $0.43^{+0.20}_{-0.20} $ & 3485/3010   &    $1.43^{+0.10}_{-0.10}$  & $0.38^{+0.20}_{-0.18}$     & 3487/3011   \\ 
214         &  2921 &  1921  & 2.0  &  $19.8^{+4.8}_{-4.0}$ & $0.87^{+0.15}_{-0.14}$ & $0.30^{+0.10}_{-0.10}$ &  3498/3010 &   $1.82^{+0.05}_{-0.05}$   
             &   $0.30^{+0.05}_{-0.10}$ & 3582.5/3011 \\
222          & 1694 & 1021 & 0.424 &  $1.8^{+0.40}_{-0.40}$ &  $1.51^{+0.19}_{-0.19}$ &  $0.12^{+0.12}_{-0.11}$ & 3477/3010 &  1.8 
                  & $<$0.15 &  3702/3012 \\
245         &     460  &  259 & 2.68 & $106.0^{+13.0}_{-30.0}$   & $1.76^{+0.82}_{-0.82}$  & $<$0.37   & 3388/3010  & $1.53^{+0.10}_{-0.28}$ 
                   & $<$0.44 & 3404/3011         \\
289          &   3790 & 2295 & 0.607 &  $1.30^{+0.27}_{-0.25}$ & $1.23^{+0.06}_{-0.12}$ & $0.18^{+0.08}_{-0.08}$ & 3330/3010 & 1.8 & 
            $<$0.10 & 4263/3012\\  
324          &  733   &  104 & 1.222 & $<0.60$ &    $0.90^{+0.30}_{-0.22}$    & $1.0^{+0.62}_{-0.54}$ & 3425/3010  &  1.8 &  $0.51^{+0.42}_{-0.35} $ 
                     &  3457/3012 \\
\hline \hline 
\end{tabular}
\begin{list}{}{}
\item The columns are:
(1) \xmm  ID.
(2) Net \xmm counts (all three modules).
(3) Net \chandra counts.   
(4) redshift; two and tree decimal digits denote photometric and spectroscopic redshift respectively. 
(5) $\rm N_H$ column density in units of $\rm 10^{22}$ \cunits. 
(6) Photon index.  (7) Equivalent-width in units of keV; the energy of the line has been fixed at a rest-frame energy 
 of 6.4 keV. (8) c-statistic value / degrees of freedom (9) Photon-index for the {\sc PEXRAV} model. (10) Equivalent-width 
 of the FeK$\alpha$ line in the {\sc PEXRAV} model. The line energy is fixed to 6.4 keV. (11) c-statistic value/degrees of freedom. All errors refer
   to the 90\% confidence level. \end{list}
\end{table*}

\section{X-ray spectral fittings and selection method} 

The goal is to identify heavily obscured AGN, via X-ray spectral analysis. 
We use the {\sc XSPEC} v12.5 software package for the spectral fits \citep{Arnaud1996}.
We employ the C-statistic technique \citep{Cash1979}, which had been specifically developed
to extract spectral information from data of low signal-to-noise ratio. This statistic works on un-binned data,
 allowing us, in principle, to use the full spectral resolution of the instruments without degrading it 
  by binning. 
   We perform our initial selection in the \xmm data both because of its high effective area 
   at high energies as well as for its good counting statistics. 
 We fit the \xmm spectra using the absorbed power-law model 
  {\sc PLCABS}  \citep{Yaqoob1997}.  The advantage of the {\sc PLCABS} model  is that it properly takes into account
Compton scattering up to column densities of $\rm N_H\sim 5\times 10^{24}$ \cunits.  

 Our selection method is  summarised in the following criteria:
 
 1) The detection of an absorption turnover  corresponding to a column density of $N_H > 1.5\times 10^{24}$ \cunits.
  For a (mildly) Compton-thick source with a column density   $\sim 10^{24}$ \cunits, 
   the absorption turnover occurs at rest-frame energies somewhat higher than $\sim$8 keV.
   This implies that even with the relatively large effective area of \xmm at high energies,  
     it is difficult  to detect the absorption turnover  for 
       marginally Compton-thick sources at low redshift. However, at higher redshifts the 
      turnover shifts progressively to low energies 
       making the identification of Compton-thick AGN more straightforward \citep{Iwasawa2012b}. 
        For example,  for a Compton-thick source with a column density of  
         $\rm N_H\sim 10^{24}$ \cunits, the absorption turnover would shift 
       to energies about 2 keV at a redshift of z=2, an energy region where \xmm has large effective area. 
   At a column density of $5\times 10^{24}$ \cunits   the turnover  occurs at 
     an energy of about 20 keV  \citep[e.g.][]{Yaqoob1997}. 
      This  criterion  is reliable only  in the case of sources 
     with a spectroscopic redshift available. This is because the rest-frame energy 
      of the absorption turnover (and hence the exact value of the column density) 
       requires knowledge of the redshift with high precision.  

 2) The detection of a flat spectral index $\Gamma < 1.4$ at a statistically significant level (90\% confidence)
  i.e. the 90\% upper limit of the photon index should not exceed $\Gamma=1.4$.    
    This is an arbitrary selected, albeit extremely conservative, cut-off. The average photon index of AGN ranges 
     between 1.7 and 2.0 with a standard deviation as low as  0.15 
   \citep[e.g.][]{Nandra1994, Dadina2008, Brightman2011,Ricci2011,Burlon2011}.   
  Therefore, a  flat spectrum may be characteristic of heavily obscured, reflection-dominated 
   Compton-thick AGN \citep{George1991, Matt2004}. 
      Note however that, because of spectral degeneracy, it is likely that some sources 
       appear to have flat spectra simply because of moderate ($\sim$ $10^{22-23}$ \cunits) absorption
        \citep[see also][]{Corral2011}. This degeneracy is pronounced in sources with low signal-to-noise spectra.
           
    3)  For the sources which are selected according to the above two criteria, we further examine whether the addition of a 
     Gaussian component is  required by the data.  
      This is additional criterion is motivated by the fact that 
       strong FeK$\alpha$ lines are often observed in Compton-thick AGN in the local Universe \citep[e.g.][]{Fukazawa2011}, 
      
    \section{The  Compton-thick candidates}
    
    \subsection{\xmm results}
The \xmm spectral fittings yield nine sources with either a flat spectrum 
 \footnote{In addition, there are two more sources which appear to have flat spectra 
  but nevertheless the 90\% upper limits of their photon indices are just above the chosen threshold of $\Gamma=1.4$.
   These are sources PID-64 and PID-252 at a spectroscopic redshift of z=0.516 and 1.893 respectively.}
 or a absorption turnover corresponding to a rest-frame column density greater than $\rm N_H  \approx 10^{24}$  \cunits. 
 The \xmm  spectral fits ({\sc PLCABS + GA} in {\sc XSPEC} notation)   
 of these nine sources are given in Table \ref{xmm}.
  Eight, out of nine candidate Compton-thick sources, have a spectroscopic redshift available. 
  One out of those (PID-245) has a noisy optical spectrum and a highly uncertain spectroscopic redshift, (1.864) 
   based on  only one line \citep{Balestra2010}. For this source there are also three photometric redshifts  
    available 2.28, 2.43 and 3.0, \citep{Dahlen2010, Santini2009, Luo2010}, respectively.
   However, it is possible that the X-ray source does not correspond to the counterpart with the available optical spectrum (located 1 arcsec away).
 \citet{Iwasawa2012b} detect an FeKalpha line which corresponds to a redshift of z=2.68. 
 We chose to use  this redshift instead of the optical ones. At this X-ray redshift there is no obvious optical lin
   
     For source PID-214, for which there are only photometric redshifts available, z=1.5 and z=1.17 \citep{Cardamone2010}, 
    there is a hint for a line at  an energy of $E=2.13^{+0.04}_{-0.03}$ keV with $\rm EW=0.3\pm0.1$ keV. 
     On the basis of this line,  assuming it is related with the FeK$\alpha$ at 6.4 keV rest-frame energy, the implied redshift  would be z=$2.00\pm0.05$. 
      Hereafter, we adopt the X-ray redshift for this source. 

  At least one source (PID-144) at a (spectroscopic) redshift of z=3.70 is a candidate transmission dominated 
  Compton-thick source i.e. is characterised as Compton-thick  
  on the basis of an  absorption turn-over  in its X-ray spectrum. This source has first been 
   reported as Compton-thick by  \citet{Norman2002}. A much higher quality X-ray spectrum  of this source
    has been reported in \citet{Comastri2011}.  
  The remaining sources present a  flat spectral index which may be 
   suggestive of a reflection continuum. 
   The FeK$\alpha$ lines provide additional information on the nature of our sources.
   Four out of nine sources (PID-66, 144, 147, 324) present high rest-frame FeK$\alpha$ EW (see table 1). 
    These are suggestive of high absorbing column densities likely corresponding to Compton-thick AGN.
  %  In Fig. \ref{66}, we present the PN spectrum of source PID-66. The \xmm spectra of the 
   %   'secure' Compton-thick sources PID-144, PID-147 have been presented in \citet{Comastri2011}.
   %   Finally, the \xmm spectrum of the source PID-324 has been presented in \citet{Georgantopoulos2011b}
%\begin{figure}
%\begin{center}
%\includegraphics[width=8.0cm]{66.pdf} 
%\caption{The \xmm PN X-ray spectrum of the candidate Compton-thick source PID-66}
%\label{66}
%\end{center}
%\end{figure}

        \subsection{Joint \xmm/\chandra spectral fits}
    
    \subsubsection{Power-law and  Fe line model}
    To increase the photon statistics, we present  
   the combined \xmm and \chandra spectral fits. These are given in Table \ref{xmm-chandra}.
    Note that, there are about 70 \xmm and \chandra 
      observations spanning a period  of more than ten years. Then, in principle one should use 
       70 different normalisations which is not feasible because of computational limitations.
        Therefore, for the sake of simplicity, the \xmm and the \chandra power-law normalizations
     have been tied to the same value.   
     
\begin{figure*}
\begin{center}
\includegraphics[width=8.0cm]{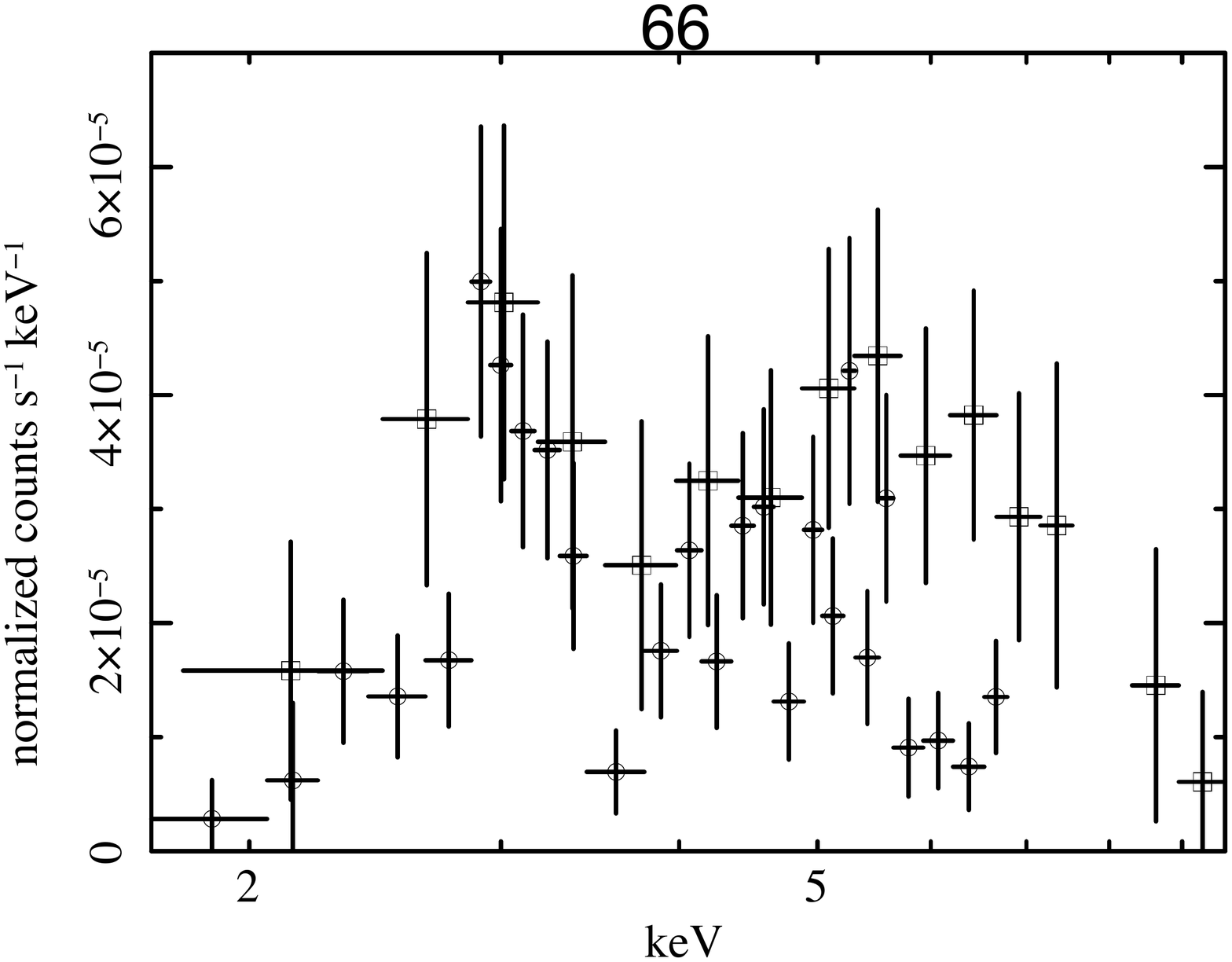} 
\includegraphics[width=8.0cm]{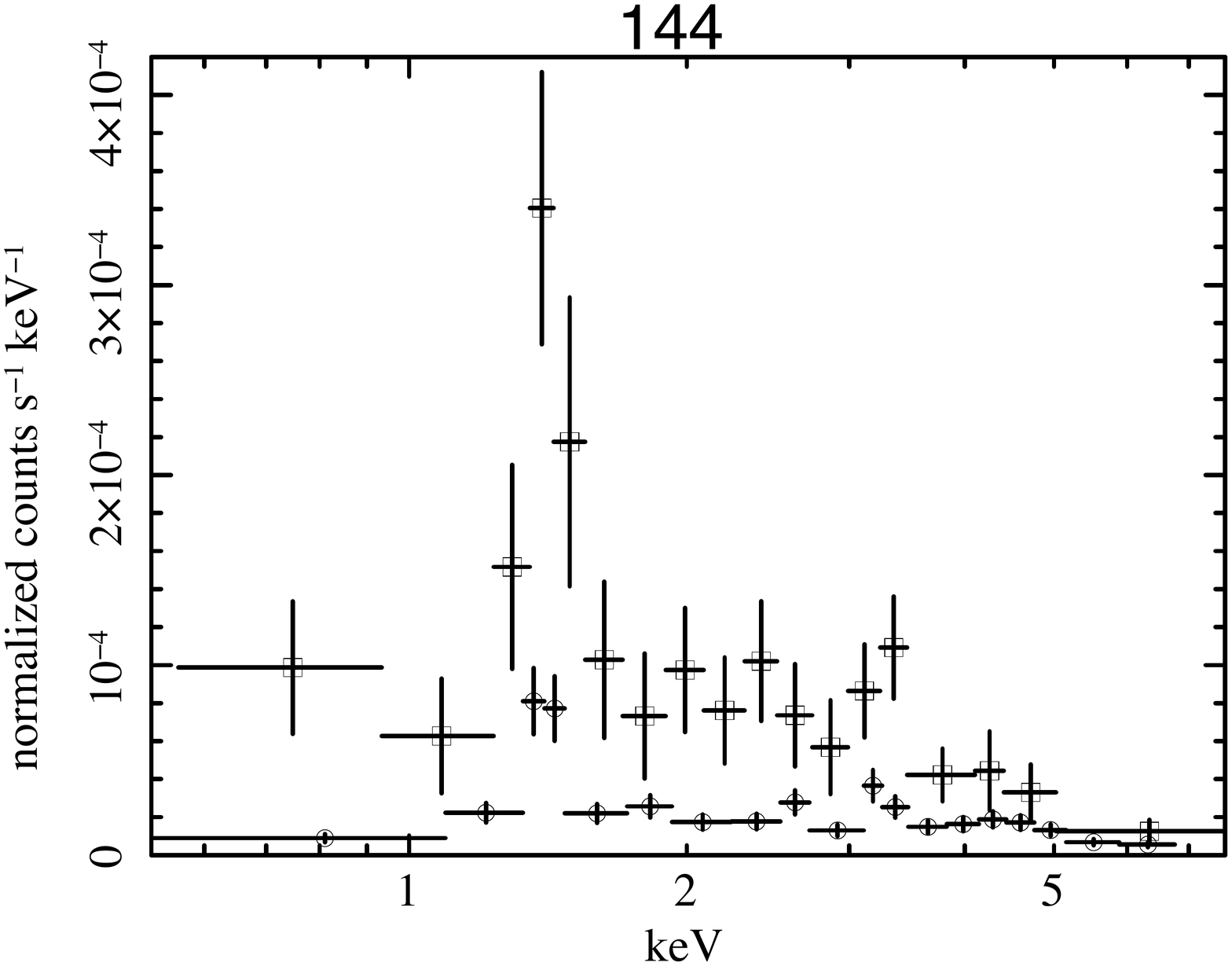} 
\includegraphics[width=8.0cm]{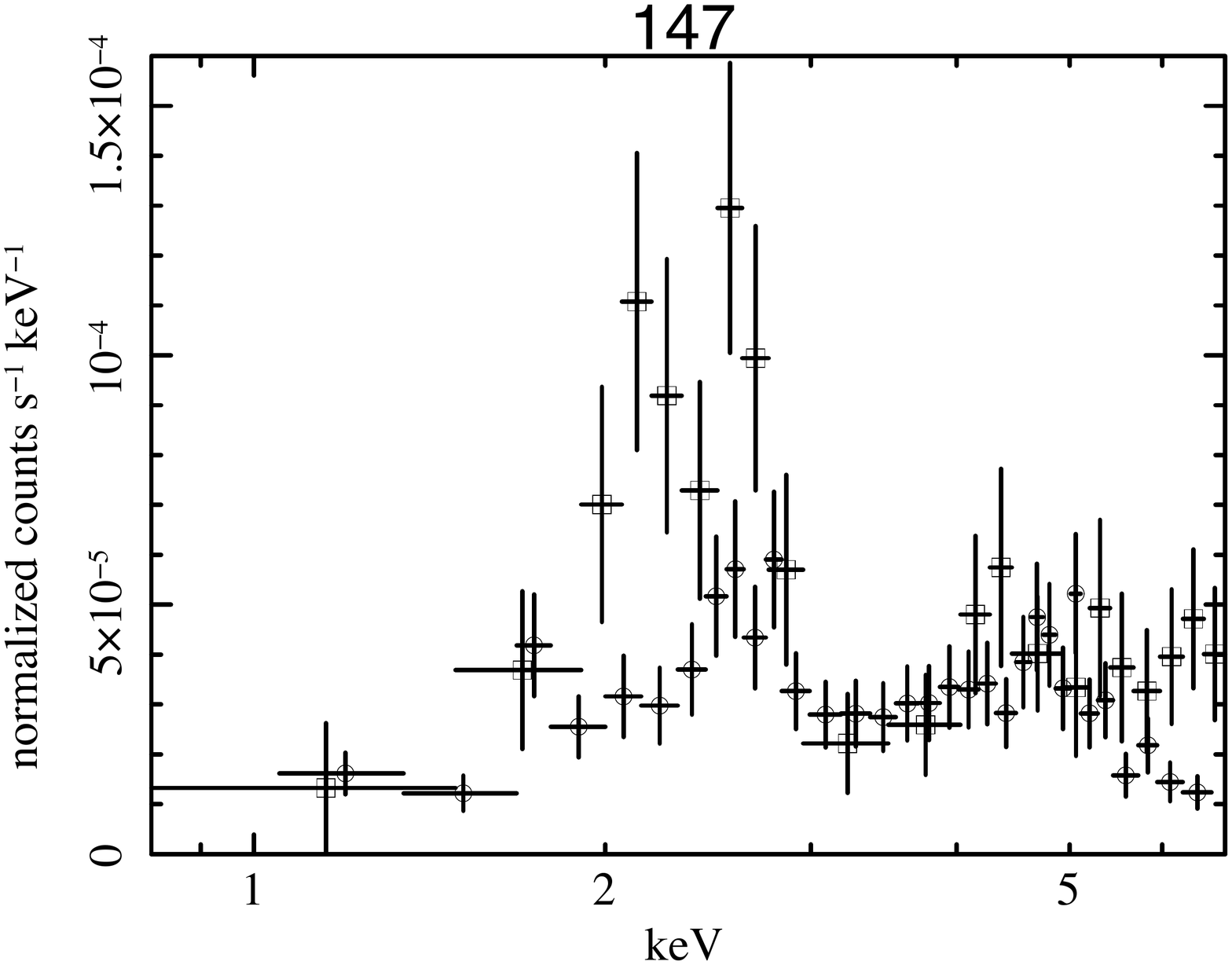} 
\includegraphics[width=8.0cm]{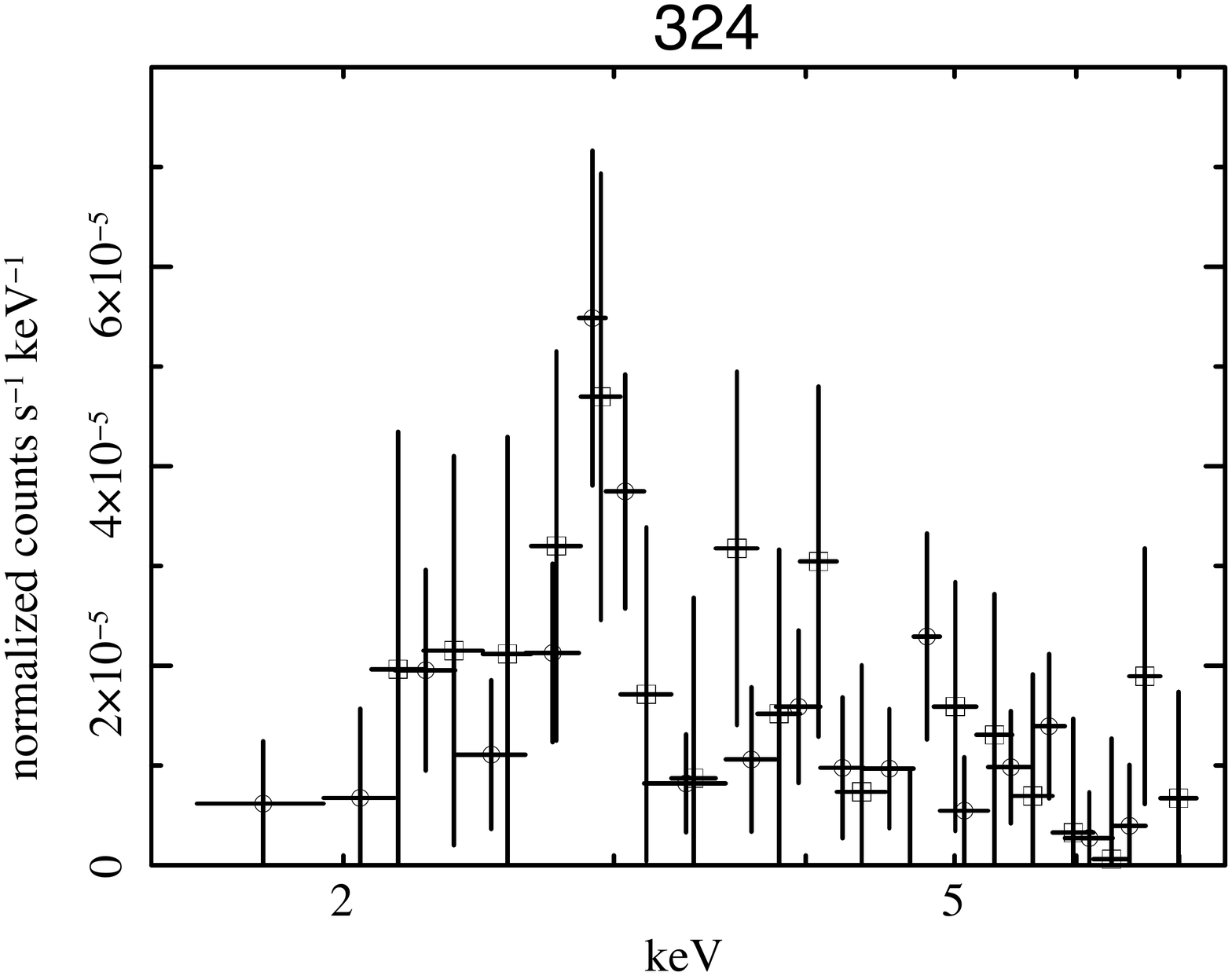} 
\caption{The X-ray \chandra (circles) and \xmm PN (squares) X-ray spectra of the candidate Compton-thick sources PID-66, PID-144 and PID-147 and PID-324.
The first and second row give the transmission and reflection dominated AGN respectively according to the combined \chandra and \xmm spectral fits. Note the 
 high spectral response of \xmm compared to \chandra at high energies.}
\label{spectra}
\end{center}
\end{figure*}

    In agreement with the \xmm spectral analysis, presented in the previous section, 
    four sources  (PID-66, 144, 147 and 324)  
   present FeK$\alpha$ lines with high (rest-frame) EW ($>$0.4 keV).
   Therefore, the joint \chandra and \xmm analysis corroborates that these four sources 
   have a good probability for being Compton-thick. 
   In Fig. \ref{spectra} we present the joint \chandra and \xmm PN spectral fits.

      \subsubsection{Reflection model} 
        Moreover, we fit a reflection model, \citet{Zdziarski1995} ({\sc pexrav + ga} in XSPEC notation). 
        The parameters in the {\sc PEXRAV} model are: the Fe abundance, the inclination angle of the 
         reflecting slab, the cut-off energy of the incident power-law and finally the slope of the
         incident power-law. We found that the first three parameters remain practically unconstrained
          owing to the large uncertainties. Therefore,  
         we fix the Fe abundance to 1, the inclination angle of to 45 degrees 
         while we assume that the incident power-law spectrum has no high energy cut-off
          \citep[see e.g.][]{Dadina2008}.

   For four sources, the derived photon index is either very flat (PID-66, $\Gamma=0.80\pm0.2$) 
   or very steep (PID-222, $\Gamma= 2.95\pm 0.08$ 
    (PID-289, $\Gamma= 2.90 \pm 0.06$),  (PID-324, $\Gamma=2.93^{+0.31}_{-0.36}$).
     As these values are many $\sigma$ discrepant with the  intrinsic spectral slopes encountered in 
      AGN \citep[see e.g][]{Nandra1994,Dadina2008}, we choose to fix the photon index to $\Gamma =1.8$. 
      The results are shown again in Table 2. 
    
    \subsection{More complex spectral models for the four 'secure' Compton-thick AGN candidates}
    For the four 'secure' candidate Compton-thick sources, we present some additional spectral models for the 
     joint \xmm and \chandra spectral fits. First, 
      we are exploring the exact energy of the FeK$\alpha$ line. This is motivated by the 
       results of \citet{Iwasawa2009} who find ionized Fe emission in many Ultra-luminous IRAS galaxies.
        Moreover, we investigate the effect of leaving the ratio between the \chandra and the \xmm 
         power-law normalization free. Note that we keep the Fe line normalization constant, as this is believed to arise 
         far away from the black hole in type-2 AGN \citep{Shu2011}. 
        Therefore, the present model has  two more free parameters: the line energy and the \chandra power-law normalization 
     which is  not tied up to the \xmm power-law normalization (Table \ref{complex}).   
     Note that the normalization of the \chandra and \xmm power-laws are consistent within the errors
     (see discussion in section 5.2.1)
      
      A further model contains an additional power-law whose  normalization has been set to 
       3\% of the  primary power-law. The slope of this power-law has been fixed to the same
        value as the primary (transmitted) power-law. This component models the scattered (unabsorbed)
         component along the line of sight which is often detected in type-2 AGN \citep[e.g.][]{Turner1997}. 
         In this model again the energy of the FeK$\alpha$ line 
         is left as a free parameter. The same holds for the normalization of the \chandra power-law which is not tied to the 
          \xmm normalization.  The results are presented in Table \ref{scattered}. The spectral fits are in general consistent 
           with the simple power-law plus FeK$\alpha$ line fits. The exception is source 66 
            where the photon-index becomes flatter at the expense of a lower column density.
             In this case, because of the flatter power-law, the EW becomes higher reaching a value of 1 keV in the case  of  \chandra.  
             In good agreement with the power-law model presented in section 5.2.1, the four sources  (PID-66, 144, 147 and 324)
             present FeK$\alpha$ line EW higher than $\sim$0.4 keV corroborating their classification as 
             good Compton-thick candidates.

 \begin{table*}
\centering
\caption{Joint \xmm and \chandra spectral fits to the four 'secure' Compton-thick sources Power-law + FeK$\alpha$ line
 ({\sl PLCABS+GA}) leaving the relative \xmm and \chandra normalization and the energy of the line free.} 
\label{complex} 
\begin{tabular}{ccccccc}
\hline\hline 
 PID       & $\rm N_H$    & $\Gamma$   &  E   &   $EW_X$    &   $EW_C$   &   c    \\
(1)               &              (2)    &         (3)       & (4)                 & (5)                & (6)   & (7)    \\
\hline
66         &    $78.0^{+4.0}_{-8.0}$ & $1.52^{+0.03}_{-0.03}$ & $6.45^{+0.25}_{-0.40}$  &  $0.23^{+0.30}_{-0.10}$ & $0.37^{+0.10}_{-0.05}$ & 3560/3008 \\
144            &  $80.0^{+7.5}_{-7.5}$ & $1.79^{+0.10}_{-0.10}$ & $6.49^{+0.46}_{-0.14}$ & $0.47^{+0.30}_{-0.24}$ & $0.56^{+0.44}_{-0.28}$ & 3475/3008 \\  
147        &  $20.0^{+4.0}_{-8.0}$  & $0.57^{+0.33}_{-0.33}$ & $6.40^{+0.13}_{-0.10}$  & $0.35^{+0.20}_{-0.15}$ &  $0.43^{+0.23}_{-0.20}$ & 3477/3008 \\
324      &  $<$0.45 & $0.76^{+0.12}_{-0.14}$  & $6.38^{+0.95}_{-0.95}$  &  $0.69^{+0.63}_{-0.42}$  & $1.15^{+0.69}_{-1.00}$ &  3402/3008 \\
\hline \hline 
\end{tabular}
\begin{list}{}{}
\item The columns are: (1) \xmm ID. 
(2) column density (in units of $10^{22}$) (3) photon index as derived from the {\sc PLCABS} model. 
 (4) FeK$\alpha$ line energy (5) \xmm rest-frame EW of the FeK$\alpha$ line. (6) \chandra rest-frame EW of the FeK$\alpha$ line. 
 (7) c-statistic value and degrees of freedom. Note: the model is the same as that used in Table 2 with the   difference that
  the FeK$\alpha$ line energy is free and also the \chandra power-law normalization is not tied to the \xmm one.    
 \end{list}
\end{table*} 

 \begin{table*}
\centering
\caption{Joint \xmm and \chandra spectral fits to the four 'secure' candidate Compton-thick candidates,  
 using a power-law + FeK$\alpha$ line + scattered emission model ({\sc PLCABS+GA+PO}), leaving the 
  relative \xmm and \chandra normalization and the central energy of the line free} 
\label{scattered} 
\begin{tabular}{ccccccc}
\hline
\hline
 PID       & $\rm N_H$    & $\Gamma$   &  E   &   $EW_X$    &   $EW_C$   &   c    \\
(1)               &              (2)    &         (3)       & (4)                 & (5)                & (6)   & (7)    \\
\hline
   66       & $28.3^{+6.1}_{-7.0}$ & $0.16^{+0.04}_{-0.07}$ & $6.45^{+0.28}_{-0.32}$ &   $0.41^{+0.25}_{-0.20}$ &  $0.99^{+0.60}_{-0.40}$ & 3551/3007  \\
  144     &  $135^{+18.0}_{-16.0}$ & $2.2^{+0.24}_{-0.32}$ & $6.5^{+0.15}_{-0.15}$ & $0.65^{+0.40}_{-0.30}$ & $0.70^{+0.40}_{-0.30}$ & 3458/3007 \\
  147     &  $8.0^{+6.0}_{-4.0}$ & $1.00^{+0.38}_{-0.31}$ & $6.57^{+0.30}_{-0.15}$ & $0.28^{+0.28}_{-0.23}$ & 
           $0.30^{+0.30}_{-0.25}$ & 3490/3007 \\
  324  & $<$0.42  &  $0.72^{+0.13}_{-0.16}$ &  $6.39^{+0.08}_{-0.08}$ & $0.51^{+1.49}_{-0.25}$ & $1.22^{+0.90}_{-0.70}$ &
        3402/3007 \\           
   \hline \hline 
\end{tabular}
\begin{list}{}{}
\item 
\item The columns are: (1) \xmm ID. 
(2) column density (in units of $10^{22}$) (3) photon index as derived from the {\sc PLCABS} model. 
 (4) FeK$\alpha$ line energy (5) \xmm rest-frame EW of the FeK$\alpha$ line. (6) \chandra rest-frame EW of the FeK$\alpha$ line. 
 (7) c-statistic value and degrees of freedom.    
 \end{list}
\end{table*}

   \section{Co-added X-ray spectra} 
   Although there are at least four sources with large FeK$\alpha$ line EW,  
   it is possible that some of the remaining sources have large EW which fail detection owing to the 
     limited photon statistics. In order to answer this question,   
  we derive the co-added X-ray spectrum. 
   The objective is to detect faint FeK$\alpha$ emission that cannot readily be detected in a single source
    \citep[e.g.][]{Iwasawa2012b}.
  The data stacking is a straight sum of the rest-frame spectra of individual sources. 
 The individual spectra are binned so that they have  a 2-10 keV rest-frame energy
range with 200 eV channel width.  The data, after background subtraction, are corrected for an efficiency curve 
 determined by the response matrix and auxilliary response file. Finally, each energy bin is corrected for redshift. 
 
\begin{figure*}
\begin{center}
\includegraphics[width=6.0cm]{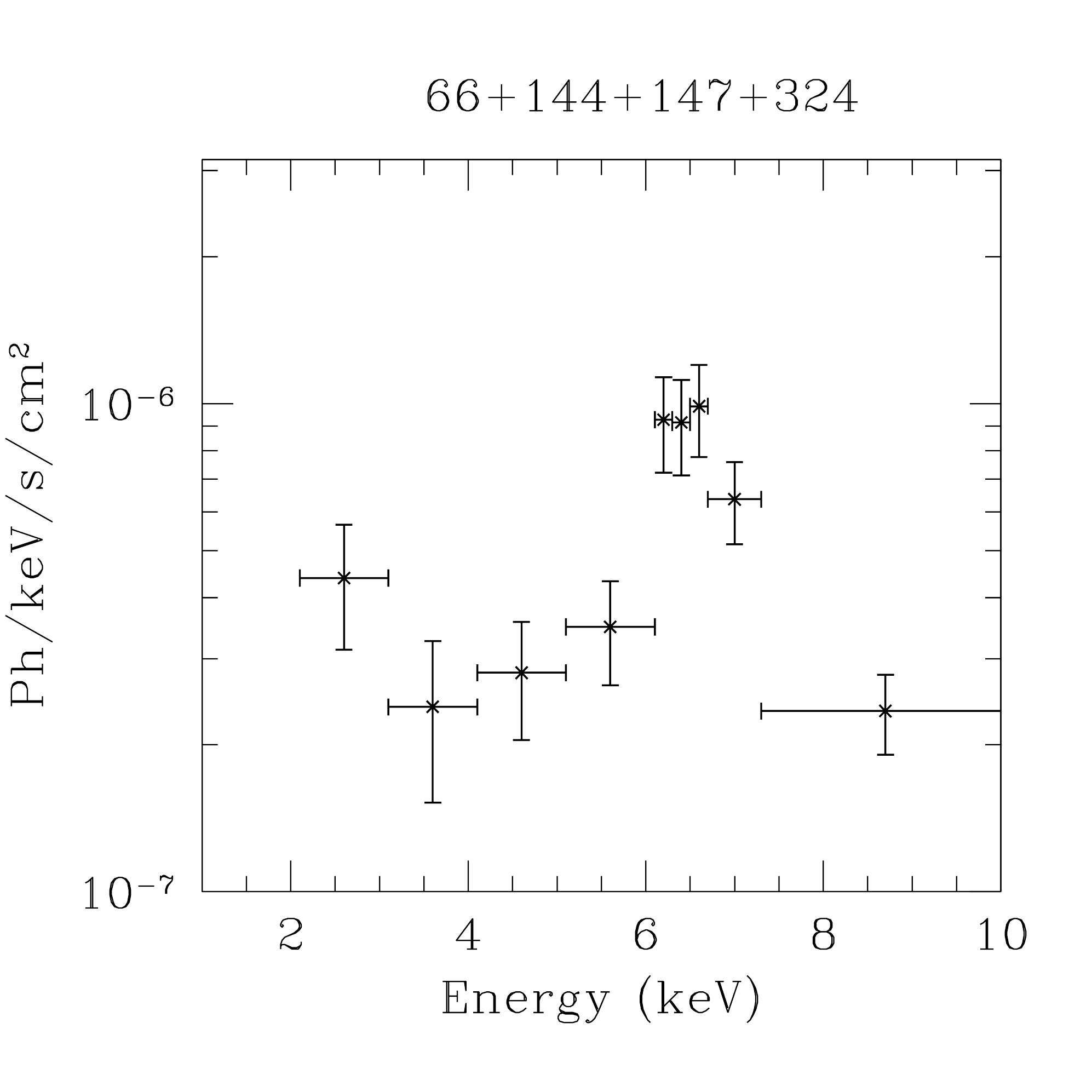}
\includegraphics[width=6.0cm]{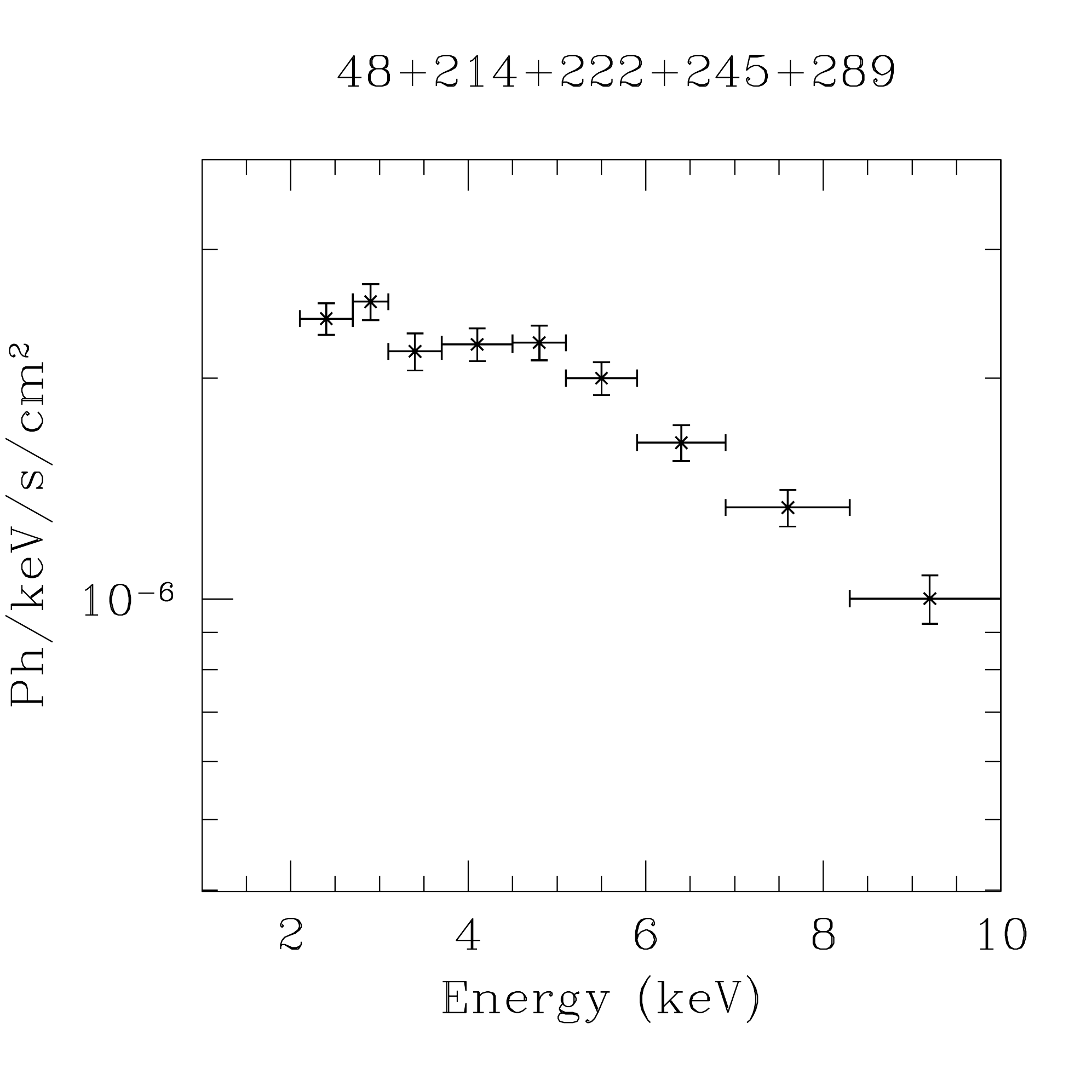}
\caption{The co-added (rest-frame) X-ray spectra for two sets of sources. Left panel: the four 
 'secure' candidate Compton-thick AGN (see text). Right panel: the remaining five sources.}
\label{coadded}
\end{center}
\end{figure*}

In Fig. \ref{coadded},  we present the co-added rest-frame X-ray spectra for two groups of sources.  
 The first group includes the four 'secure' Compton-thick AGN candidates (PID-66, 144, 147 and 324). 
   The second group includes the remaining sources (PID-48, 214, 222, 245 and 289). 
    A difficulty with the second group is that two of the redshifts (PID-214 and 245)
     are based on the X-ray spectra and thus may be more ambiguous.  
      The first  group of sources displays a very prominent FeK$\alpha$ line at a rest-frame
       energy of 6.4 keV. The second group possibly shows an emission feature 
      but at a higher energy of about 7 keV. If confirmed, this could imply the presence of an ionized FeK$\alpha$ line.

 \begin{figure*}
\begin{center}
\includegraphics[width=5.0cm]{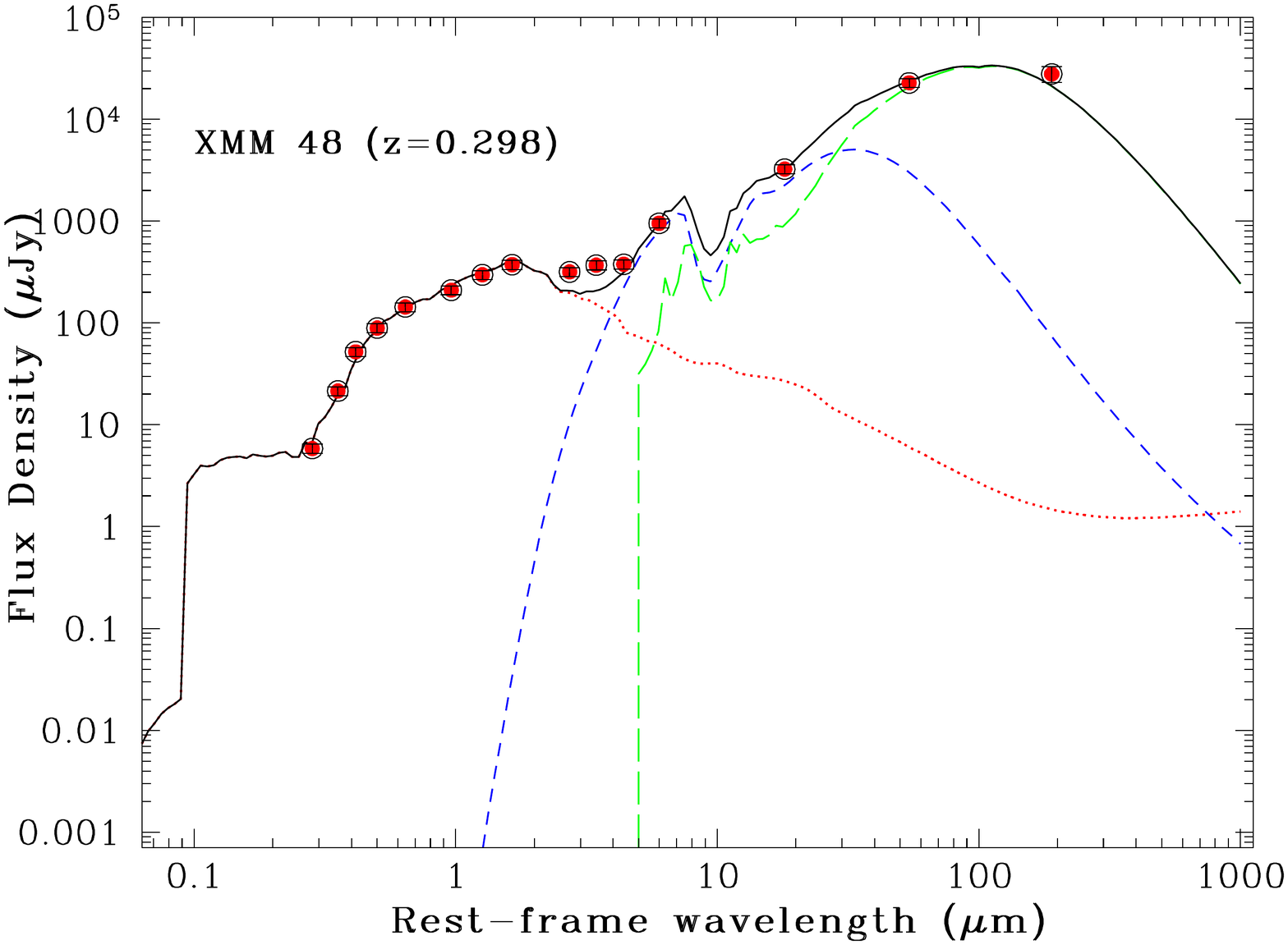} 
\includegraphics[width=5.0cm]{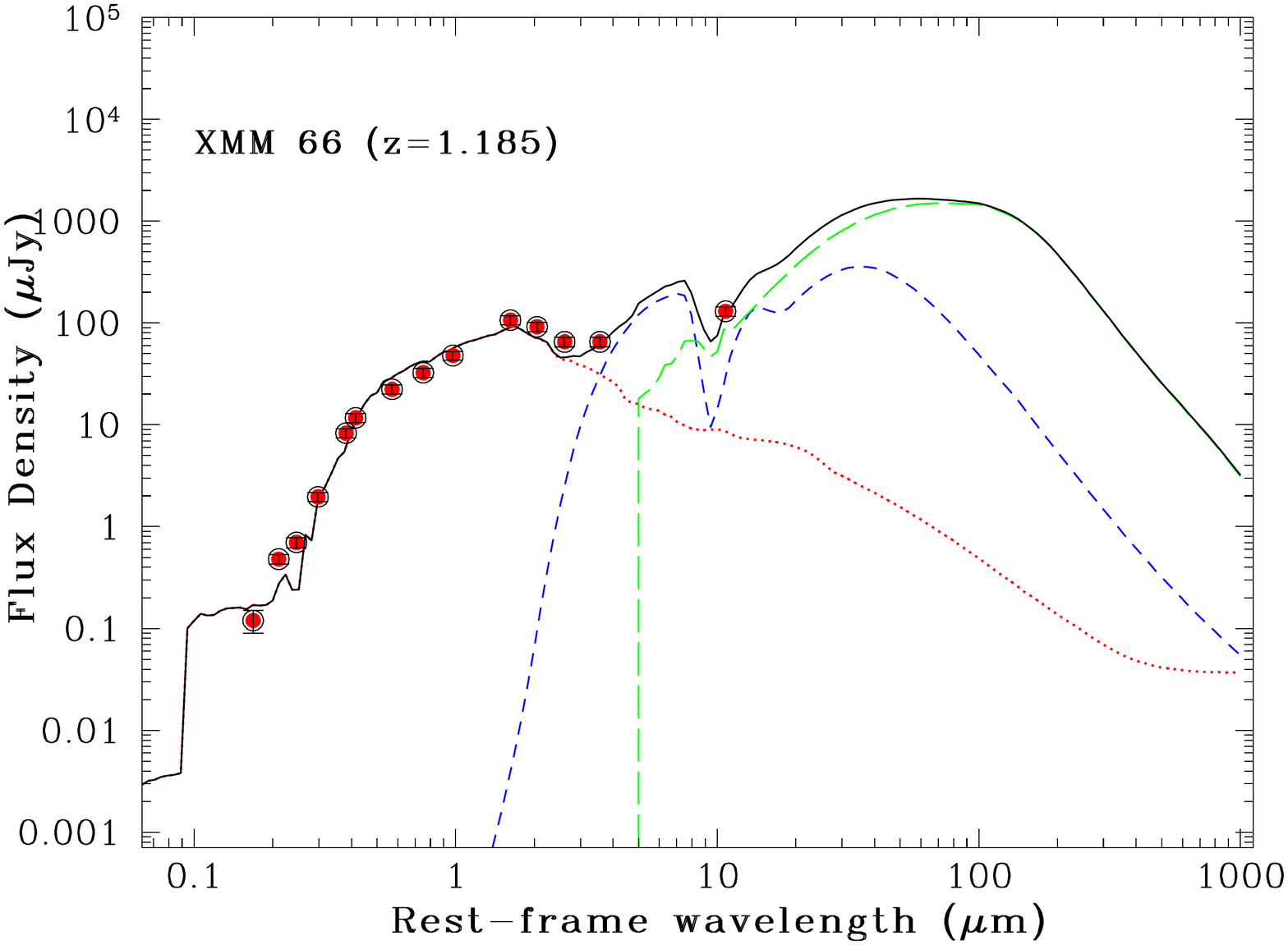}
\includegraphics[width=5.0cm]{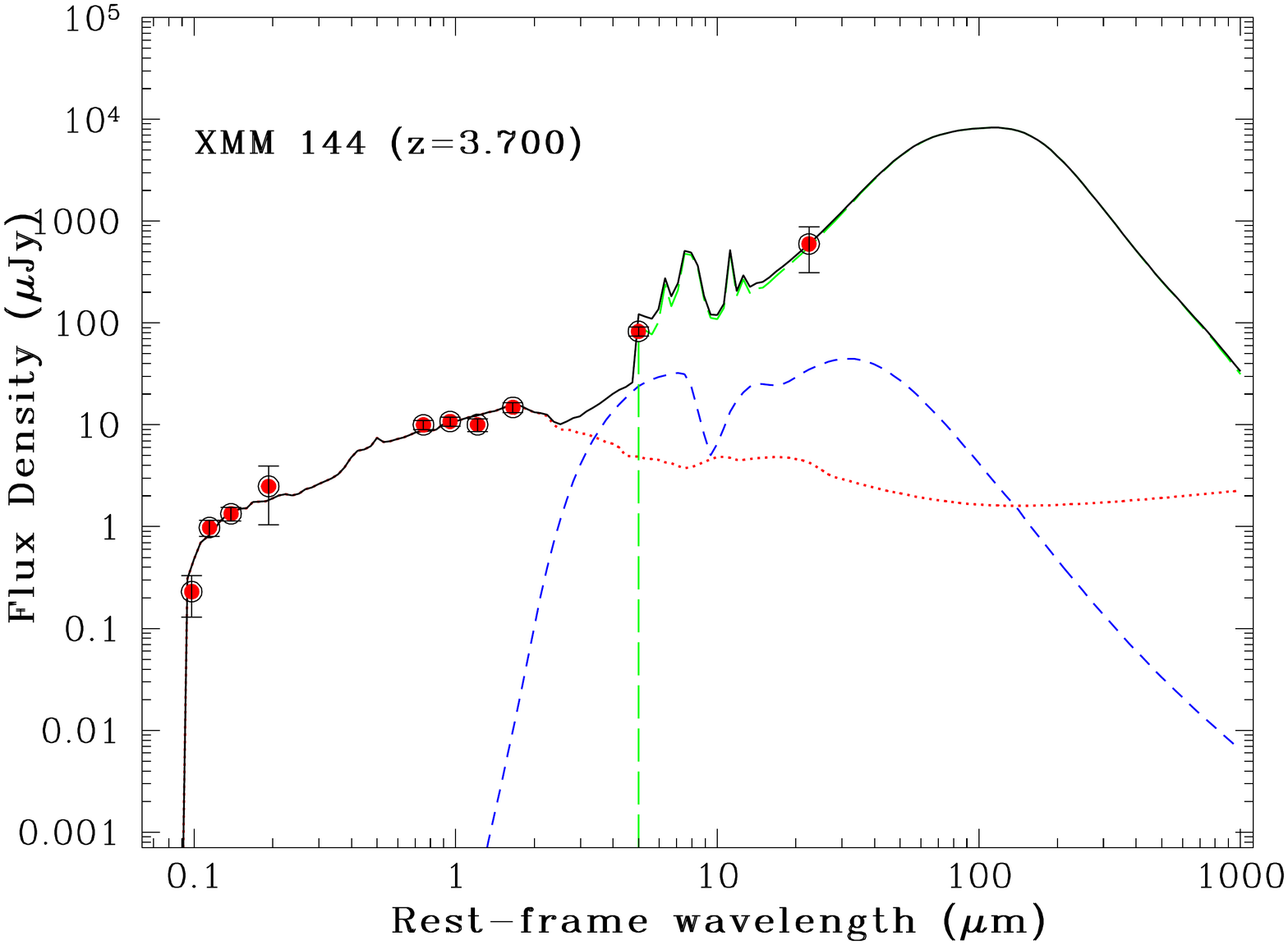} \vfill
\includegraphics[width=5.0cm]{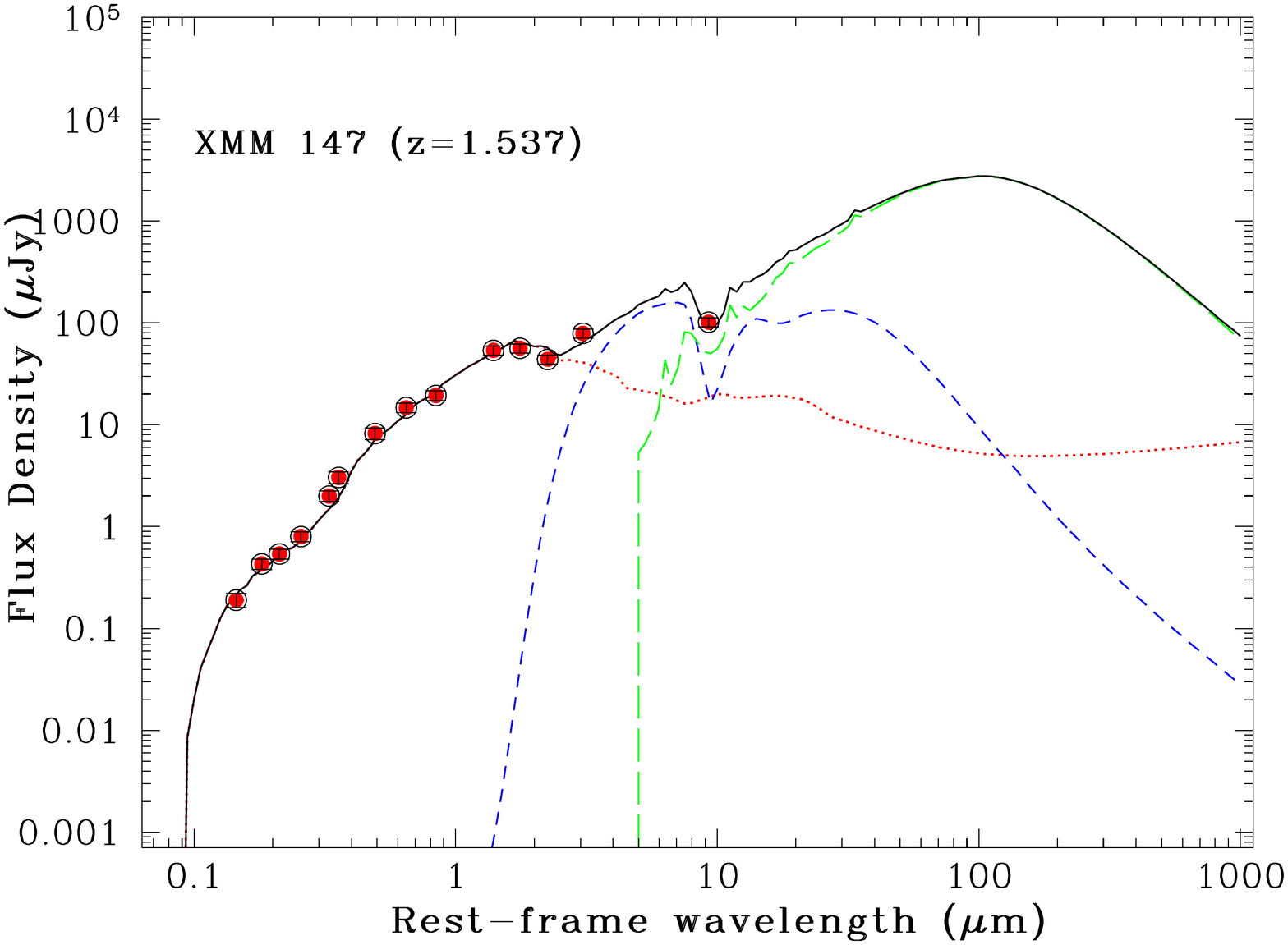}
\includegraphics[width=5.0cm]{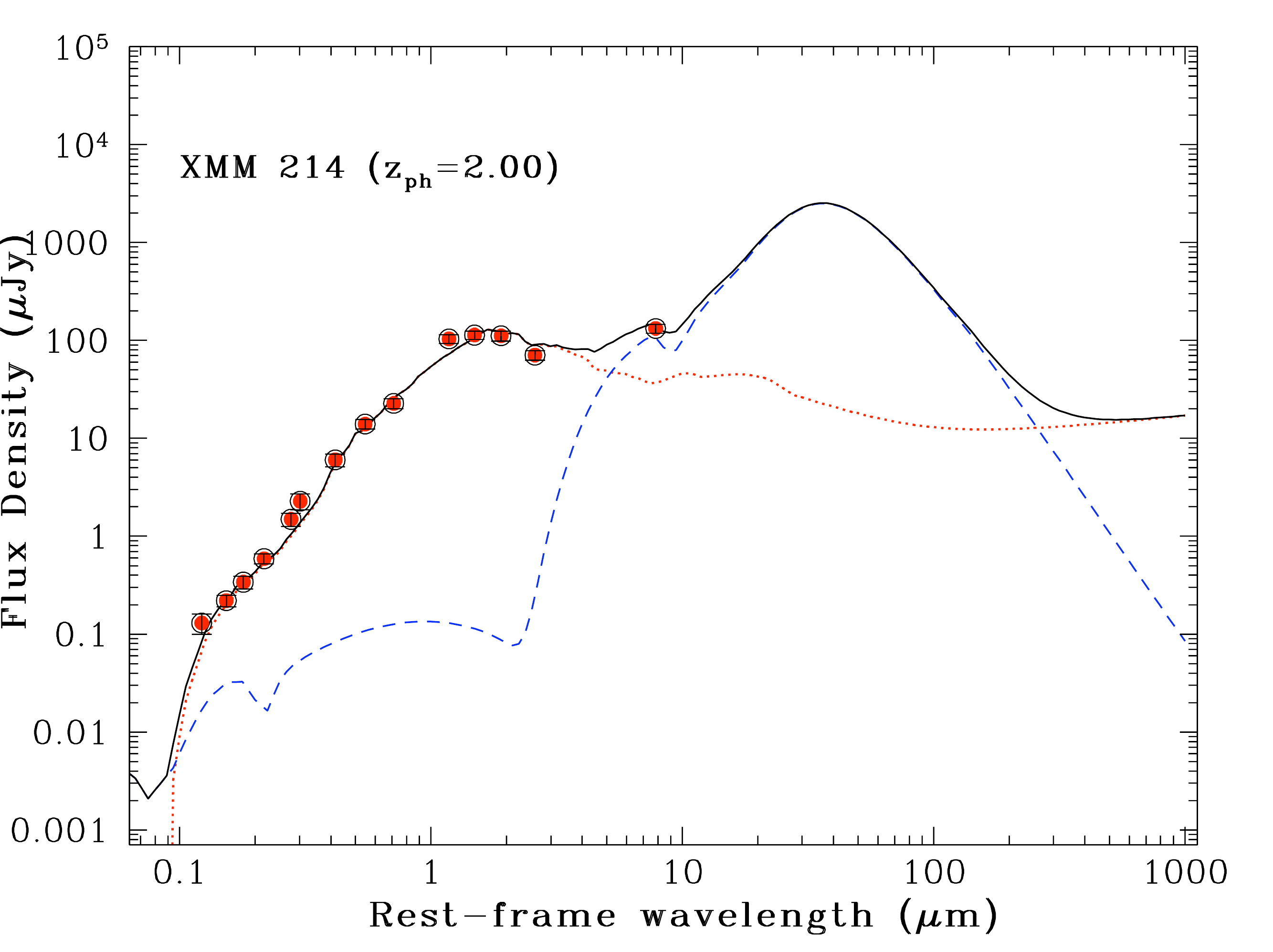}
\includegraphics[width=5.0cm]{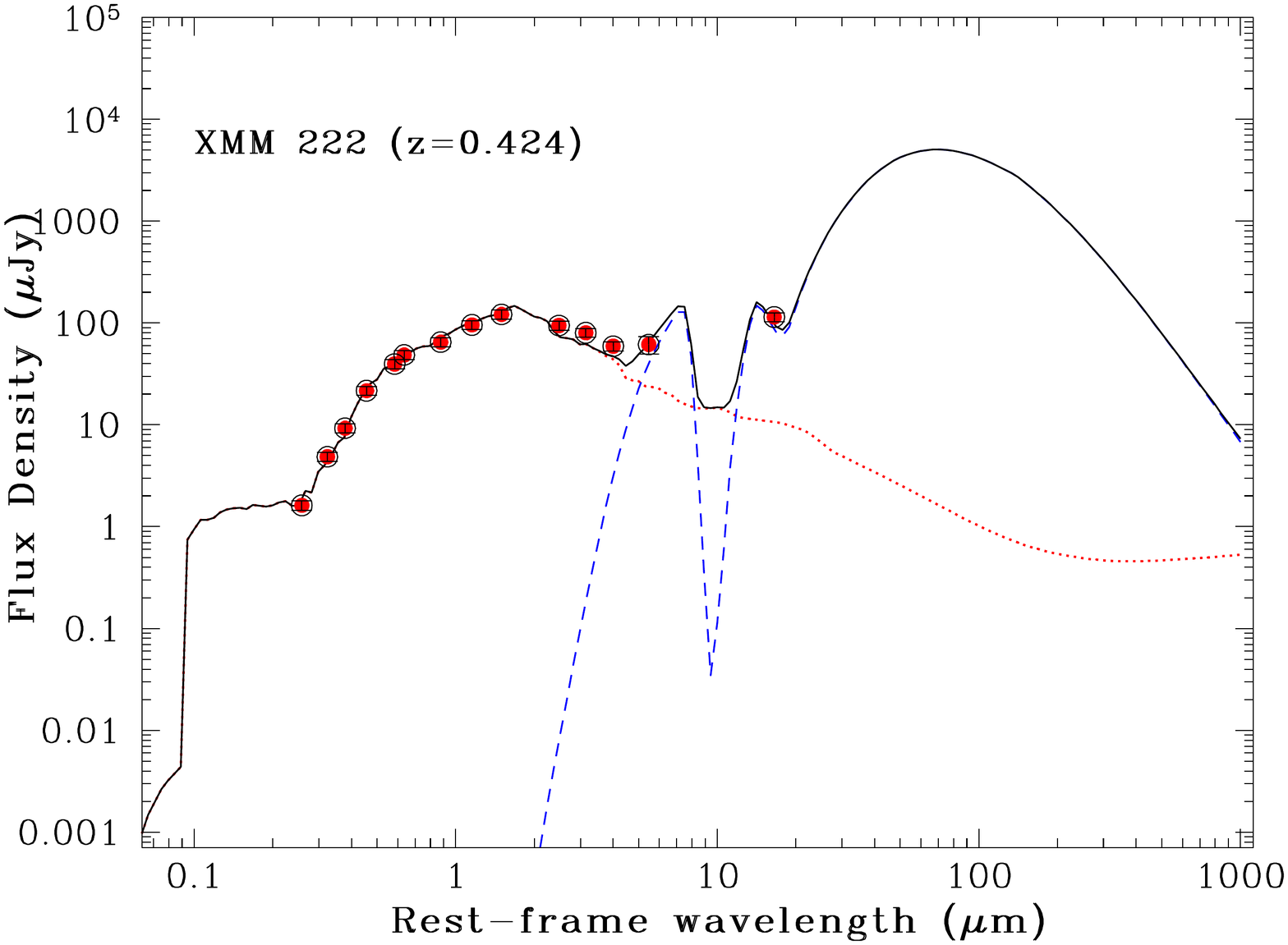} \vfill
\includegraphics[width=5.0cm]{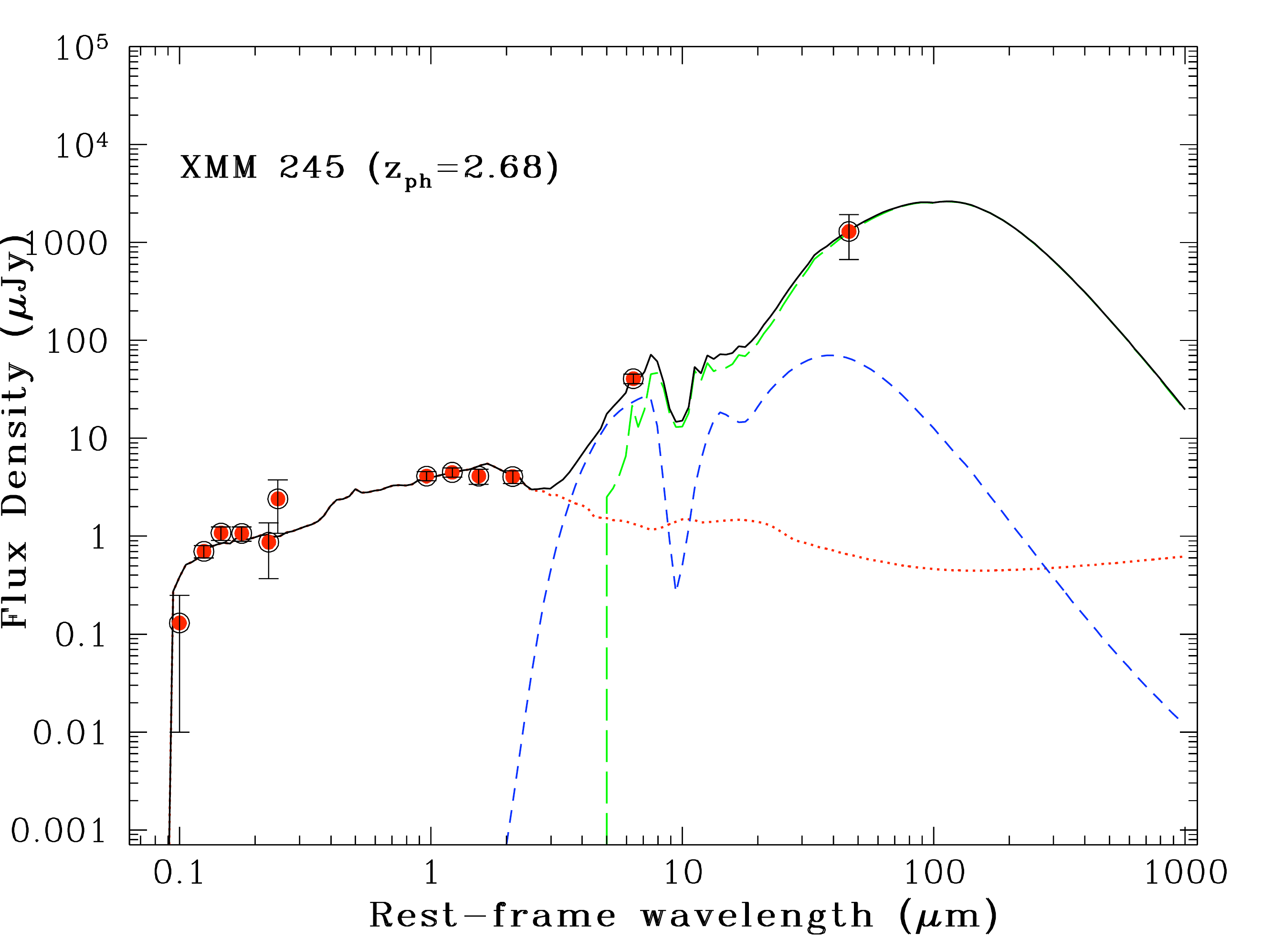}
\includegraphics[width=5.0cm]{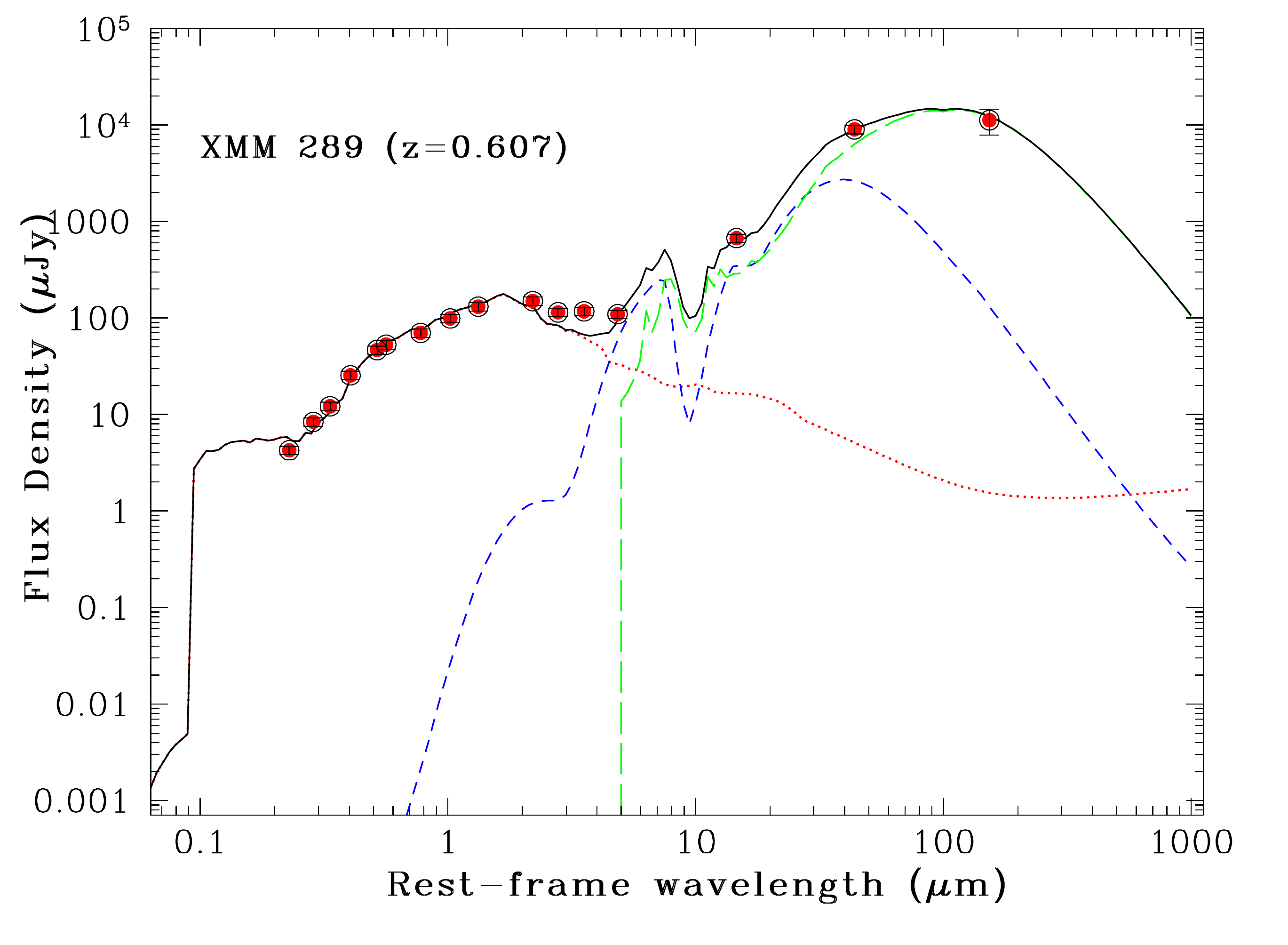}
\includegraphics[width=5.0cm]{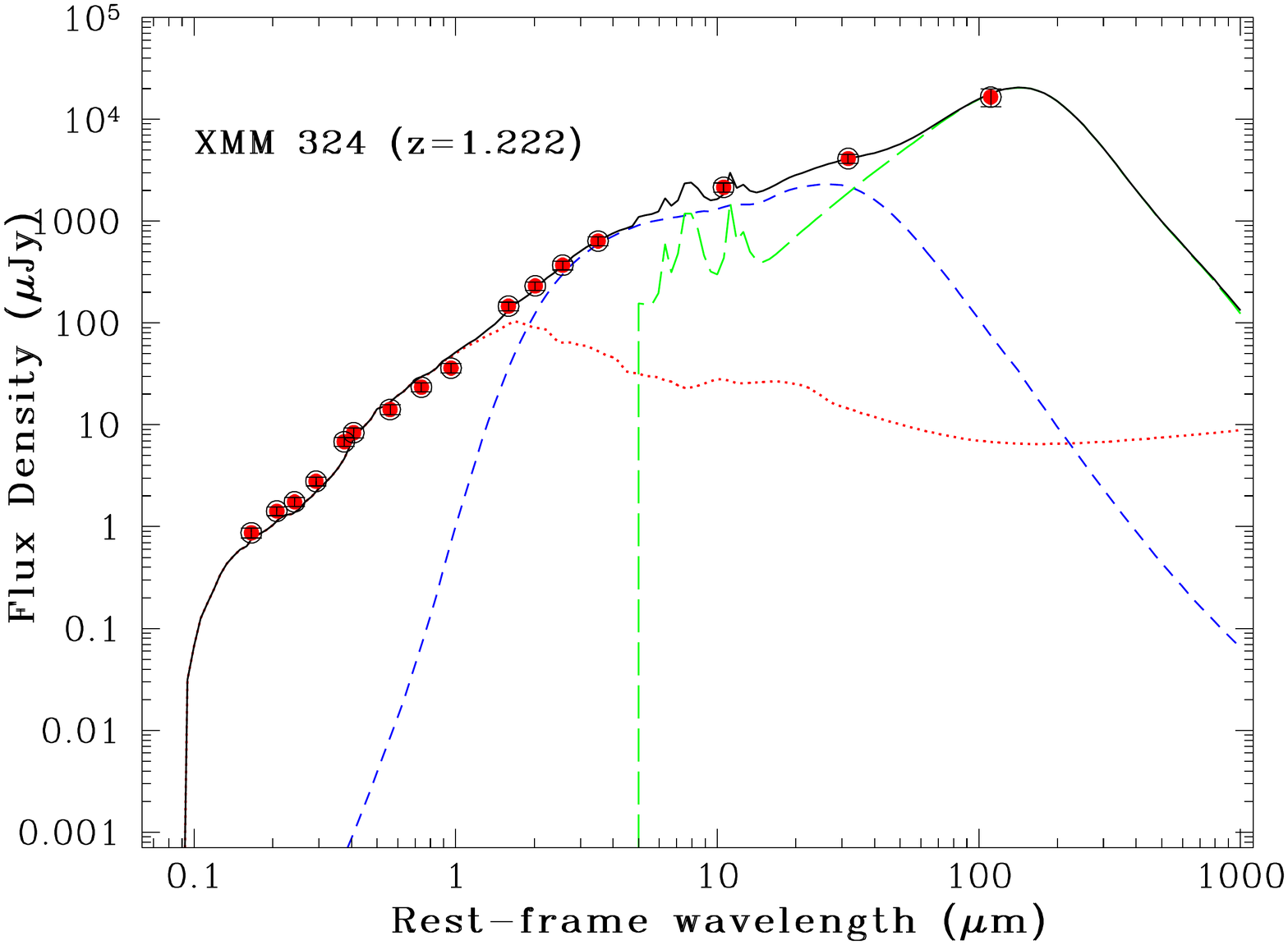}  \vfill
\caption{The spectral energy distributions of the nine Compton-thick candidate sources. The red (dot), blue (short dash) and green (long dash) 
 lines denote the stellar, torus and starburst component respectively, while the black line denotes the sum of the three 
  components.}
\label{sed}
\end{center}
\end{figure*}
       
\section{IR properties}

\subsection{Spectral Energy Distributions}

We construct SEDs (Spectral Energy Distribution) for the full sample 
 of our flat spectrum sources, to  constrain the
predominant powering mechanism (AGN or star formation) in the mid/far IR part of the
spectrum. The SEDs are also used to obtain an accurate estimate of the
$\rm 12\mu m$ nuclear infrared luminosities and thus the X-ray to mid-IR luminosity ratios 
 which are often considered as good diagnostics for heavily obscured sources. 

 We use data from UV wavelengths to the far-IR (where available). 
Four out of nine sources do not have a far-IR measurement available from {\it Herschel}.   
The data have been modelled using the code originally developed by \citet{Fritz2006}.
  The code has been updated by \citet{Feltre2012}. 
The SED fitting is based on a multicomponent analysis, e.g. \citet{Vignali2009}, \citet{Hatziminaoglou2009} \citet{Pozzi2012}.
The observed UV to far-IR SED is de-composed in three distinct components: a) stars, having the bulk of the emission 
 in the optical/near-IR 
b) hot dust, mainly heated by UV/optical emission due to gas accreting on to the supermassive black hole 
 c) colder dust heated by star formation.
The stellar component has been included using a set of simple stellar population (SSP) spectra of solar metallicity and ages ranging from $\approx$1 Myr to $\approx$8 Gyr.
  A common value of extinction is applied to stars of all ages, and a \citet{Calzetti2009} attenuation law has been adopted 
   ($\rm R_V= 4.05$).
   In order to model the emission at observed wavelengths greater than 24$\mu m$, the SED fitting also includes a component from colder, diffuse dust, likely heated 
    by start formation processes, as discussed below.
 This  is represented by templates of known starburst galaxies \citep{Vignali2009}.
 The SED fits are shown in Fig. \ref{sed}. The four 'secure' candidate Compton-thick sources appear to have 
  star-forming components. However, the strength of the star-forming component in PID-66 and PID-147 is ambiguous because of the 
   lack of far-IR data in these sources. 
   Note that the use of templates of nearby starbursts to model the far-IR emission 
    and hence star-formation rates, may present some limitations. 
    This is because there may be discrepancies between typical IR SEDs at high redshift compared to 
    local sources \citep[e.g.][]{Elbaz2010, Nordon2012}. 
   
   Next, we derive the exact star-formation rates and 
   specific star-formation rates, i.e. the ratio of star-formation over the stellar mass.      
            The far-IR luminosity is probably  the most reliable tracer of star-forming activity
          \citep{Kennicutt2012}.    The IR photons are emitted by the dust surrounding young stars,
which is heated by their ultra-violet radiation.
          The star-formation rate is derived from the total IR luminosity ($\rm 8-1000 \mu m$) which is ascribed to star-formation according to 
           the SED decomposition.  We use the relation between the IR luminosity and SFR in \citet{Murphy2011}:
 \begin{equation}
 \rm SFR [M_\odot yr^{-1}] = 3.88 \times 10^{-44} L(8-1000 \mu m) [erg s^{-1}]
 \end{equation}     
  Then, the specific star-formation rate, sSFR,  is given by
  \begin{equation} 
          \rm logsSFR [Gyr^{-1}]=logSFR [M_\odot yr^{-1}] - log M^{\star} [M_\odot] + 8.77
          \end{equation}
           The stellar mass, $M^{\star}$ is estimated from the optical part of the SED   \citep[for details see][]{Rovilos2012}. 
           Table \ref{irtable} summarises the IR luminosities, star-formation rate and stellar masses for 
          the heavily obscured AGN in our sample. The typical errors on the 12$\rm \mu m$ are 
         about 20\% apart from PID-48, PID-66, PID-222 and PID-245 where 
           the uncertainties can be as high as a factor of two. In most cases, the typical errors 
     on the total IR luminosity are between 10 and 20\%. In the cases where no far-IR data 
     are available, the errors become large (factors of 2-3). Finally, for the stellar masses the errors 
      are of the order of 10\%.

  \begin{figure*}
\begin{center}
\includegraphics[width=13.0cm]{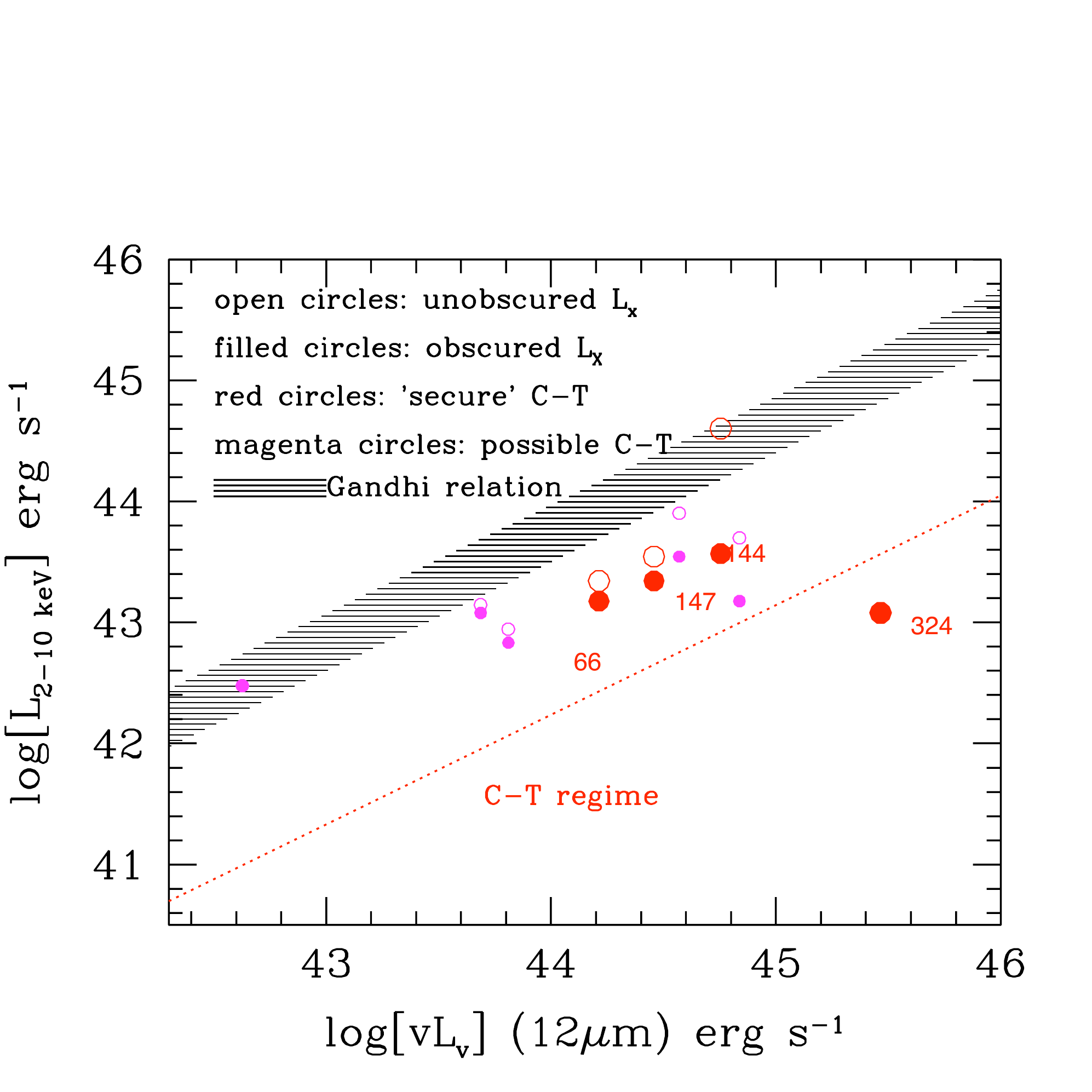} 
\caption{The absorbed X-ray (2-10 keV) X-ray luminosity against the 12$\rm \mu m$ luminosity.  The large circles 
 correspond to the best candidate Compton-thick sources as suggested by the large equivalent-width FeK$\alpha$  lines.
  The typical errors are of the order of 30\% and 20\% for the IR and X-ray luminosity respectively, 
   including the uncertainties in the model fitting. 
 The hatched diagram represents the $1\sigma$ envelope of the local  \citep{Gandhi2009} relation. 
  Open circles correspond to X-ray luminosities
  corrected for absorption (see table 1). For two sources (PID-222 and PID-324)
   the absorbed luminosity equals the unabsorbed luminosity. 
  The dotted line corresponds to a factor of 30 lower X-ray luminosity as is typical in many Compton-thick nuclei.}
\label{lxl12}
\end{center}
\end{figure*}

% \begin{figure*}
%\begin{center}
%\includegraphics[width=3.8cm]{cdf153.pdf} 
%\includegraphics[width=3.8cm]{cdf156.pdf} 
%\includegraphics[width=3.8cm]{cdf202.pdf} 
%\includegraphics[width=3.8cm]{cdf268a.pdf} 
%\caption{The optical spectra of three candidate Compton-thick sources 
% from \citet{Szokoly2004}. The notation follows that of \citet{Giacconi2002}:
 % CDFS-153 (XMMID-147), CDFS-156 (XMMID-66), CDFS-268a (XMMID-324). On the horizontal axis the rest-frame (top) and observed 
 %  (bottom) wavelength in $\rm \AA$ units is shown.}
%\label{optspec}
%\end{center}
%\end{figure*}

\begin{figure*}
\begin{center}
\includegraphics[width=10.0cm]{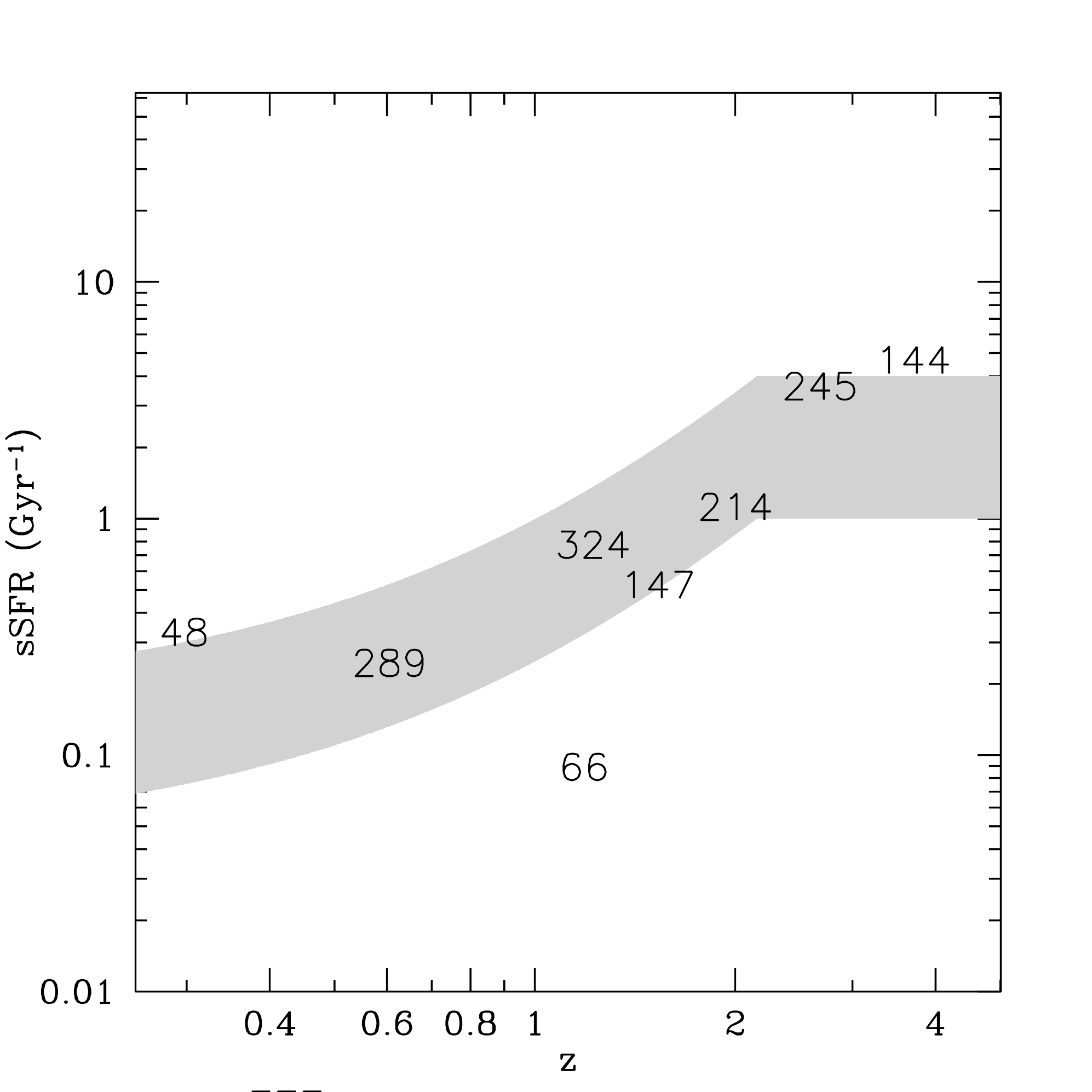} 
\caption{The specific star-formation rate as a function of redshift for eight of the candidate Compton-thick sources.
The shaded area gives the star-formation main-sequence according to \citet{Elbaz2011}. Above and below  
 this area lie the starburst and quiescent galaxies respectively. Source 222 is not ploted as there is no available rest-frame photometry 
  above a rest-frame wavelength 24$\rm \mu m$. }
\label{ssfr}
\end{center}
\end{figure*}

\begin{table*}
\centering
\caption{Infrared properties of heavily obscured sources} 
\label{irtable} 
\begin{tabular}{cccccccc}
\hline\hline 
 PID              &  z &    $L_{12}$ & $L_{8-1000}$ & AGN frac. & $logM^{\star}$ & SFR  & sSFR  \\
(1)               &              (2)    &         (3)       & (4)                 & (5)             & (6) & (7)  & (8)    \\
\hline
48      & 0.297  &   1.66 &  1.05 & 0.25 &  10.51 & 1.26  & -0.48 \\
66      & 1.185   & 4.21 &   1.40  &  0.22  & 11.21 & 1.39  &  -1.05  \\
144    & 3.700   &  14.60 &  33.7 &  0.01 &  10.87 & 2.77 &  0.67  \\
147    &  1.530   & 7.37 & 3.33 & 0.10  &  10.81 & 1.76  &  -0.28 \\ 
214     & 2.00  &  9.5 & 6.5  &  0.95  &  10.52 & 1.90 &  -0.99 \\ 
222      & 0.424 &  0.11 &  -$^{\dagger}$   & -   & 10.64  &   - &  -  \\
245       &  2.68 &  17.7 &  6.6  &  0.05 &  10.34 & 2.05  &  0.56 \\
289       & 0.605 &   1.25 & 2.0 & 0.22  &  10.81 & 1.43  &  -0.61  \\
324        & 1.222 &  75.3 &  8.3 & 0.36  &  10.84 & 1.96 &  -0.11 \\
\hline \hline 
\end{tabular}
\begin{list}{}{}
\item The columns are:
(1) XMM-ID ; (2) redshift (3) rest-frame 12$\rm \mu m$ luminosity in units of solar  luminosities ($\times 10^{10}$) ; (4) total IR 
 luminosity in units of solar luminosities  ($\times 10^{11}$); (5) AGN fraction of the total IR luminosity (6) logarithm of stellar mass in units of
  solar masses (7) logarithm of star-formation rate 
  in units of solar masses/yr;    (8) specific star-formation rate in units of Gyr$^{-1}$.     $^{\dagger}$: no  photometry above 24$\rm \mu m$ rest-frame.  
\end{list}
\end{table*}

\subsection{X-ray to IR luminosity ratios} 
                  
     The detection of a low X-ray to mid-IR luminosity ratio has  been widely used as the
main instrument for the detection of faint Compton-thick AGN which cannot be easily identified 
 in X-ray wavelengths  \citep[e.g.][]{Goulding2011}. This is
because the mid-IR luminosity (e.g. $\rm 12\,\mu m$ or $\rm 6 \mu m$)  is a good proxy of the AGN power,
as it should be dominated by very hot dust which is heated by the AGN
\citep[e.g.][]{Lutz2004,Maiolino2007}. At these wavelengths the contribution
by the stellar-light and colder dust heated by young stars should be small. 
\citet{Gandhi2009} present high angular resolution mid-IR ($12 \mu m$) observations of the nuclei
 of 42 nearby Seyfert galaxies.  These observations provide the least contaminated 
core fluxes of AGN. These authors find a tight correlation between the near-IR fluxes and the 
 intrinsic X-ray luminosity. 
   Although the {\it Spitzer} observations do not have the spatial resolution to resolve the core,
    the SED decomposition allows us to derive with reasonable accuracy the nuclear IR luminosity.  
  In Fig. \ref{lxl12}, we present the obscured X-ray luminosities against the 12 $\rm \mu m$ 
   luminosities.  All four candidate Compton-thick sources have low $\rm L_X/L_{12}$  ratios.
 Therefore, it appears that the use of the (obscured) X-ray to mid-IR (12$\rm \mu m$) ratio is a reliable diagnostic
  for the presence of heavily obscured sources. Note that the unobscured X-ray luminosity of at least 
  two sources lies below the  \citet{Gandhi2009} relation. This may suggest that these are Compton-thick 
   despite the absence of high-EW FeK$\alpha$ lines.

\subsection{Star formation}

 It has been suggested that highly obscured AGN  at X-ray wavelengths  may be associated with large amounts of 
        star-formation \citep[see for example][]{Georgakakis2003, Rovilos2007, Mainieri2011}. 
        \citet{Alexander2005} find that many sub-mm galaxies, which have very high rates of star-formation, are 
         associated with heavily absorbed sources  in X-ray wavelengths. 
        A possible explanation for such a link  would be that the nuclear star-forming region is 
         associated with the X-ray absorbing screen.  From Fig. \ref{sed}, it appears that in the four  'secure' Compton-thick sources, 
          the rest-frame mid-IR wavelengths (around 6$\mu m$) are dominated by the torus emission. 
          Only two of the four 'secure' Compton-thick sources have long-wavelength data  (PID-144, PID-324), and 
           the star-forming component is dominated in both cases. 
             In Fig. \ref{ssfr} we plot the specific star-formation rate  as a function of redshift.              
             The four candidate Compton-thick sources have a
            specific star-formation which roughly  follows the  main star-forming sequence as described by \citet{Elbaz2011}.  
             Moreover, it appears that the star-formation properties of the Compton-thick AGN are not different from those of 
              X-ray selected AGN in general \citep[see e.g.][]{Mullaney2011,Santini2012, Rosario2012, Rovilos2012}.  
              Therefore, it appears that there is no link between the presence of a Compton-thick 
               nucleus and enhanced star-formation activity in our sample.

\section{Discussion}

\subsection{Limitations in the sample selection and completeness} 
We compile a sample of  candidate Compton-thick AGN searching for either a) flat spectrum 
 sources  i.e. those with $\Gamma < 1.4$ (at the 90\% confidence level)
  which is suggestive of a reflection dominated spectrum  \citep[e.g.][]{George1991} 
  or b) those showing an absorption turnover suggesting a rest-frame column density 
   of $>10^{24}$ \cunits. 
   The deep flux limits of the present observations facilitates the detection of Compton-thick sources. 
   This is because Compton-thick AGN are difficult to detect because they are very faint in the X-rays for the same intrinsic luminosity, 
    meaning that many Compton-thick sources at a given luminosity will be missed because of the effective survey flux limits. 
   
   We find nine candidate sources 
    among the 176 \xmm spectra examined.
    Out of these,  four show FeK$\alpha$ lines with large equivalent-widths
     and hence these have higher probability of being Compton-thick.   
    Our sample is by no means complete. For example, at faint fluxes there may be 
    sources which have flat spectra, albeit with larger error bars on $\Gamma$
     and hence fail our selection criterion. 
     Incompleteness may also be introduced by the errors in uncertain redshifts. 
      This is important in the cases where the Compton-thick classification is based 
       on the detection of an absorption turnover. 

Next, we discuss whether there are apparently flat X-ray spectrum sources which are not genuinely 
  reflection dominated Compton-thick sources.  
  AGN have a steep power-law photon index with a slope of $\Gamma\sim1.8$ and a dispersion 
   of $\sigma=0.15$ \citep[e.g.][]{Nandra1994}.
    Sources which have a significantly flatter photon-index have a large probability of being Compton-thick. 
  However, there are flat-spectrum sources which are probably not Compton-thick. 
       This is because there is a degeneracy between the photon-index and the column density
        and therefore a moderate column density or a complex/multiple absorber can be mistaken as a flat photon index.   
        For example, in table 1, we see that source PID-66 (z=1.185) shows a photon-index of 
         $\Gamma=-0.37^{+1.2}_{-0.6}$  and $\rm N_H=12^{+45}_{-11}$ \cunits (3015/2498) when both the photon-index and the column density are left free. 
          When the photon-index is fixed to the more realistic value of $\Gamma=1.8$,
           the implied column density becomes $\rm N_H\approx  100^{+20}_{-20}\times10^{22}$ \cunits, while 
            the difference in the Cash statistic (3018/2500) is not statistically significant.  
        This degeneracy can be seen even in sources with large numbers of photon counts. 
        \citet{Corral2011} present an X-ray spectral analysis of 
        $\sim$300 AGN from the \xmm  bright survey. 
         There are several sources (e.g. XBSJ134656.7+580315 at a redshift of z=0.373), at bright fluxes 
          ($\rm f_{2-10}>10^{-13}$ \funits), and hence with excellent photon statistics which present a flat power-law as the best-fit. 
           \citet{Corral2011} point out that there are no iron lines with large EW detected in these sources. 
           If the photon-index is fixed at $\Gamma=1.9$ the resulting column density is of the order 
            $\sim10^{23}$ \cunits and therefore these sources, although certainly heavily obscured, 
              are not Compton-thick.

\subsection{The X-ray  spectrum of Compton-thick AGN} 
Additional ambiguities in the selection may be introduced by 
 uncertainties in the X-ray spectrum of 
  Compton-thick AGN.  \citet{Brightman2012} perform a selection 
   of Compton-thick sources among the \chandra sources in the 4Ms CDF-S data 
    using their own spectral models. These take into account Compton-scattering, 
     the geometry of the circumnuclear matter as well as the scattered nuclear light. 
  They find a  number of sources with a large scattering component (up to 44\%)
   and hence steeper spectra. The amount of scattered light increases with increasing column density. 
    These amounts of scattered light are well above those detected in local Compton-thick AGN 
     \citep[e.g.][]{Comastri2010}. 
    Evidently,  Compton-thick sources with steep  spectra  would have avoided detection in our selection criterion. 

The strength of the FeK$\alpha$  line introduces an additional uncertainty.  
 The presence of an FeK$\alpha$ line with a large EW is considered 
  to be the 'smoking gun' for the presence of a Compton-thick nucleus.
   \citet{Comastri2010} present {\it Suzaku} observations of a few Compton-thick sources. 
    They find narrow, FeK$\alpha$ lines (EW$\sim$1-2 keV) due to neutral or mildly ionized gas  in all of them. 
   \citet{Fukazawa2011} present {\it Suzaku} observations of a sample 
    of nearby Seyfert galaxies among which many are Compton-thick. 
    They find that all Compton-thick sources present Fe$K\alpha$ lines with an EW exceeding 0.5 keV. 
   However, there have been rare examples of Compton-thick 
    sources with small EW. One of these examples is 
     the Broad-absorption-Line QSO Mrk231. Its {\it BeppoSAX} spectrum \citep{Braito2004} shows that it is 
   Compton-thick  (with  a column density of $\sim 10^{24}$ \cunits), but it has an FeK$\alpha$ line with an 
     EW of only 0.3keV. Moreover, recent {\it Suzaku} observations showed a
     decrease in the covering fraction of the absorber \citep{Piconcelli2012}.

\subsection{Previous studies of heavily obscured AGN in the CDFS} 
\citet{Tozzi2006} first derived a sample of candidate Compton-thick
sources in the 1Ms CDF-S observations based on X-ray spectral fits.
They derived a sample of 20 candidate Compton-thick sources.
\citet{Georgantopoulos2007} performed the same exercise compiling a
sample of 18 sources in the same dataset but using slightly different
spectral models. Only nine of the sources are common in these two
samples, demonstrating that the uncertainties in this technique are
considerable.  From the candidate heavily obscured sources in
\citet{Tozzi2006} or \citet{Georgantopoulos2007}, only two sources are
found in our present heavily obscured \xmm sample. 

\citet{Comastri2011} used the 3Ms \xmm data in the CDF-S attempting to
confirm that some of the Compton-thick candidates in the above samples
are indeed secure Compton-thick AGN. The excellent quality \xmm data and in
particular its ability to detect the FeK$\alpha$ line allowed the
above authors to confirm the best two examples of Compton-thick
candidates in the high redshift Universe: PID-144 and PID-147 at
redshifts of z=1.53 and z=3.7 respectively.  Note that PID-144 has
been first reported as Compton-thick by \citet{Norman2002} on the
basis of 1Ms \chandra spectroscopy.  

For PID-144, \citet{Comastri2011} fit an absorbed power-law model
finding a photon-index of $\Gamma=1.48^{+0.33}_{-0.40}$ with a column density of
$\rm 6^{+3}_{-2}\times 10^{23}$ \cunits i.e. favouring a transmission-dominated
heavily obscured source. Moreover, they detect a strong FeK$\alpha$
line with an EW of $840^{+290}_{-420}$ eV.  Our joint \xmm/\chandra
spectral fit (table 2)  suggests a transmission-dominated highly obscured
spectrum with $\rm N_H\sim 5^{+4}_{-3}\times 10^{23}$ \cunits and a photon index
of $\Gamma=1.26^{+0.72}_{-0.47}$ fully consistent with the results of
\citet{Comastri2011}.  The FeK$\alpha$ line EW is somewhat lower
($610^{+430}_{-300}$) eV but fully consistent with the above results.

In the case of PID-147, \citet{Comastri2011} find that a good fit to
the data is provided by an absorbed power-law model with a column
density of $5^{+4}_{-2}\times 10^{22}$ \cunits and a photon index of
$\Gamma=-0.11\pm 0.22$; the iron line has a very large EW
($1870^{+410}_{-450}$ eV).  Our combined \xmm and \chandra spectral
fits (Table 2) give a higher column density of $19^{+5}_{-7}\times10^{22}$ \cunits but
with a steeper photon-index $\Gamma\approx0.46^{+0.45}_{-0.20}$, from
which a smaller equivalent width of $430^{+200}_{-200}$ eV
results. Further exploration of the spectra suggests that the
difference can be ascribed to degeneracies in the parameter space,
coupled with the different treatment of statistic (C-stat vs. 
$\chi^2$) and binning of the spectra, which may result in different
local minima for the statistic parameter. 
               
  \citet{Brightman2012} 
 report  20 secure Compton-thick AGN. Out of these, five have been reported as Compton-thick in 
  \citet{Tozzi2006} while three in \citet{Georgantopoulos2007}.
   The two Compton-thick sources reported by \citet{Comastri2011} (144 and 147) and in our paper 
    have column densities just below $10^{24}$ \cunits in the spectral fits of \citet{Brightman2012}. 
    Recently, \citet{Iwasawa2012} performed a search for heavily obscured AGN in the same \xmm sample with this presented here.
     They are  using an X-ray colour-colour diagram based on the rest-frame 3-5 keV, 5-9 keV 
      and 9-20 keV bands.  They find a sample of seven high-redshift (z$>1.7$) candidate heavily obscured sources 
      having rest-frame 9-20 keV excess emission. Some of them are good Compton-thick candidates 
       either on the basis of a high-EW ($> $1 keV) FeK$\alpha$ line (e.g. PID-114 at z=1.806) or directly
        on the basis  of large   column densities, $\rm N_H\approx 10^{24}$, (PID-245, 252).  
           
             \subsection{Comparison with X-ray background synthesis models}
      Next, we  compare with  the predictions of X-ray background synthesis models
       \citep[e.g][]{Gilli2007,Treister2009,Ballantyne2011,Akylas2012}
       In our sample, we find 
       four candidate Compton-thick sources two of which appear to 
        be transmission-dominated. 
        Comparing with the models of \citet{Gilli2007}\footnote{www.bo.astro.it/$\sim$gilli/counts.html}, 
        we find that the number of Compton-thick sources 
                is  $\approx$7 in our field down to the flux limit of 
          the present survey $f_{(2-10)}\sim1\times10^{-15}$ \funits. 
          The model of \citet{Treister2009}\footnote{agn.astroudec.cl} predicts a number of 8 Compton-thick 
           in the same area. Finally, the model of \citet{Akylas2012} \footnote{indra.astro.no.gr}
          yields about  11  Compton-thick sources in our field. Note that all the above estimates 
           refer only to transmission dominated ($\rm logN_H=24-25$) Compton-thick AGN. 
           If we include the reflection-dominated AGN ($\rm logN_H=25-26$) in the model of \citet{Gilli2007}
            the total number of Compton-thick AGN rises to  $\approx11$. 
           Interestingly, the model of \citet{Akylas2012} has a lower intrinsic fraction of Compton-thick 
            AGN compared to that of \citet{Gilli2007}: 15\% compared to 33\%. 
             However, the higher number of Compton-thick AGN predicted by \citet{Akylas2012} 
             at the relatively bright fluxes probed here, is a consequence 
             of the stronger reflection component predicted by this model.
             The fact that the number of the Compton-thick candidates found here 
              is systematically lower than those predicted by all 
              X-ray background synthesis models may suggest that our sample 
               suffers from incompleteness.

\subsection{Concluding remarks}
The selection of Compton-thick AGN on the basis of either 
an absorption turnover (transmission dominated) or a flat spectrum 
(reflection dominated) appears to be sufficiently robust. 
 The presence of a high EW FeK$\alpha$ line can be 
  considered as the final diagnostic for the presence of a Compton-thick 
   AGN despite some possible exceptions mentioned above (e.g. Mrk231).
    On the basis of the above diagnostic we isolate four sources 
     as excellent candidates for being Compton-thick. These are probably the 
      best Compton-thick candidates at moderate to high redshifts found so far.
      Still, the number found here is a factor of at least two lower compared to the 
       predictions of X-ray background synthesis models.
       This may imply that a number of Compton-thick AGN (particularly 
        those at the faintest fluxes) fail to be classified as Compton-thick.
        It is likely that these  lie among our remaining five flat spectrum sources.
         Then  we did not classify them as 'secure' Compton-thick because 
          we failed to detect strong EW FeK$\alpha$ lines. 
        The  fact that a few sources lie below the \citet{Gandhi2009} 
         (unabsorbed) X-ray to IR luminosity relation may point 
          towards this scenario. This stresses the need for complementary
           IR methods together with X-ray spectroscopy, in order to better 
            understand the properties of Compton-thick AGN.      

   \section{Summary} 
   We report on an X-ray spectral study of the 176 brightest sources in the \xmm survey in the CDFS. 
    The aim is to identify very highly   obscured (Compton-thick) AGN. Our methodology consists of looking for 
        sources which have either a) an absorption   cut-off characteristic of a high column density  or b) a flat spectrum 
         (the 90\% upper limit of  the photon index should be lower than 1.4).
           After the selection of our candidate sources, we are additionally looking for the presence of 
             a strong  FeK$\alpha$ line which is considered to be the trademark of a highly obscured source.     
         Our results can be briefly summarised as follows:
     
     \begin{itemize}
     \item{The \xmm spectra select nine candidate 
      heavily obscured sources. We separate a group 
       of four sources which have 
       large iron line EW and therefore these have 
        higher probability of being Compton-thick.
         Adding the \chandra data to the spectral fits 
         corroborates our previous results. Two of the sources are 
        most likely transmission dominated (PID-66 and PID-144)
         while the other two are reflection dominated (PID-147 and PID-324)}. 
        \item{Although our sample is by no means statistically 
          complete, it represents the most well studied examples
           of Compton-thick AGN at moderate to high redshifts.
           In particular, the redshifts of the four most secure candidate Compton-thick sources 
            are z=1.185, 1.222, 1.53 and 3.7.
           One of our candidate Compton-thick sources (PID-66) 
            is presented here for the first time. 
                   The \xmm only spectra of two sources (PID-144 and PID-147) have been reported 
          in \citet{Comastri2011} while their combined XMM/Chandra spectrum is 
           presented here for the first time. Finally, one source 
            (PID-324) has been reported in detail  in \citet{Georgantopoulos2011}}.
              \item{The X-ray to mid-IR (12$\mu m$) luminosity ratio of the four candidate Compton-thick 
              are well below the average relation of \citet{Gandhi2009}, independently suggesting heavy obscuration}. 
             \item{{\it Spitzer} and {\it Herschel} observations  of our four candidate Compton-thick 
              sources derive star-formation rates between about  25 and 1000 $M_\odot~yr^{-1}$.
              Their specific star-formation rates are consistent with those of normal galaxies suggesting 
               that heavy obscuration is not related to enhanced star-formation}.  
        \end{itemize}

\begin{acknowledgements}
IG and AC acknowledge the Marie Curie fellowship FP7-PEOPLE-IEF-2008 Prop.
235285. We acknowledge financial contribution from the agreement 
 ASI-INAF I/009/10/0. PR acknowledges the receipt of a fellowship (proposal no. P9-3493) from the 
  Greek Secretariat of Research and Technology in the framework of the 
   project "support to postdoctoral researchers".  
 NC acknowledges financial support from the Della Riccia foundation. 
 FJC acknowledges partial financial support by the Spanish Ministry of Economy
  and competitiveness through the grant AYA2010-21490-C02-01.  
The Chandra data used were taken from the Chandra Data Archive at the Chandra X-ray
Center. 
\end{acknowledgements}

\end{document}